\documentclass[twocolumn]{aa} 
\usepackage[]{natbib}
\usepackage{graphicx,color}
\usepackage{txfonts}
\usepackage[colorlinks=true, citecolor=blue]{hyperref}
\usepackage{multicol}
\usepackage{amsmath}
\usepackage{graphicx}
\usepackage{tablefootnote}
\usepackage{tabularx}
\usepackage{multirow}
\usepackage{float}
\usepackage{verbatim}
\usepackage{gensymb}
\usepackage{lineno}

\makeatother

\newcommand{\orcid}[1]{\href{https://orcid.org/#1}{\includegraphics[width=10pt]{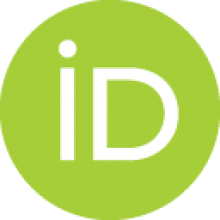}}}

\hypersetup{draft}
\begin{document} 

\title{CAPOS: The bulge Cluster APOgee Survey V. Elemental abundances of the bulge globular cluster HP~1}                       
\author{
Lady Henao\inst{1,2,3,4}\orcid{0000-0002-2036-2944}\thanks{lhenaoo@udec.cl}
\and
Sandro Villanova\inst{5}
\and
Doug Geisler \inst{1,6}
\and
Jos\'e G. Fern\'andez-Trincado\inst{7}\orcid{0000-0003-3526-5052}\thanks{jose.fernandez@ucn.cl}
}

\authorrunning{Lady Henao et al.} 
    
\institute{
        Departamento de Astronom\'ia, Universidad de Concepci\'on, Casilla 160-C, Concepcion, Chile
        \and
        Universidad Sergio Arboleda, Departamento de Matem\'aticas, Escuela de Ciencias Exactas e Ingenier\'ias, calle 74 \# 14-14, Bogot\'a, Colombia.
        \and 
        Observatorio Astron\'omico Nacional, Universidad Nacional de Colombia, Bogot\'a, Colombia
        \and 
        Departamento de F\'isica, Facultad de Ciencias, Pontificia Universidad Javeriana, Cr. 7 No 40 - 62, Bogot\'a D.C. 110231, Colombia
        \and
        Universidad Andres Bello, Facultad de Ciencias Exactas, Departamento de Ciencias F\'isicas - Instituto de Astrof\'isica, Autopista Concepcion-Talcahuano 7100, Talcahuano, Chile.
        \and 
        Department of Astronomy, Facultad de Ciencias, Universidad de La Serena, Av. Juan Cisternas 1200, La Serena, Chile
        \and
        Instituto de Astronom\'ia, Universidad Cat\'olica del Norte, Av. Angamos 0610, Antofagasta, Chile
    }
	
	\date{Received ...; Accepted ...}
	\titlerunning{CAPOS HP~1}

  \abstract  
	{ We have performed a detailed abundance analysis of 10 red giant members of the heavily obscured bulge globular cluster HP~1 using high-resolution, high S/N  near-infrared spectra collected with the Apache Point Observatory Galactic Evolution Experiment II survey (APOGEE-2), obtained as part of the bulge Cluster APOgee Survey (CAPOS). We investigate chemical abundances for a variety of species including the light (C,N), odd-Z (Al), $\alpha$  (O,Mg,Si,S,Ca and Ti), Fe-peak (Ni,Fe), and neutron-capture (Ce) elements. The derived mean cluster metallicity is [Fe/H]$=-1.15\pm0.03$, with no evidence for an intrinsic metallicity spread. HP~1 exhibits a typical $\alpha$-enrichment that follows the trend for similar metallicity Galactic GCs, such as  NGC~288 and NGC~5904, although our [Si/Fe] abundances are relatively high. We find a significant nitrogen spread ($\sim 1$ dex), and a large fraction of nitrogen-enhanced ([N/Fe]$>+0.7$) stars that populate the cluster. We also detect intrinsic star-to-star spreads in [C/Fe], [O/Fe], [Al/Fe], and [Ca/Fe], which are (anti)correlated with several chemical species, indicating the prevalence of the multiple-population phenomenon in HP 1.
    We have uncovered for the first time a possible correlation between Ca and Al, although the sample is small.
    The mean $\langle$[Mg/Fe]$\rangle = +0.29$ and $\langle$[Al/Fe]$\rangle = +0.46$ place HP~1 into the region dominated by \textit{in-situ} GCs, supporting the \textit{in-situ} nature of this cluster.
    }


\maketitle
\section{Introduction}
Globular Clusters (GCs) are fossil relics and key to help unlock our Galaxy's  formation and early evolution, since they are very old objects, formed around the time of galaxy formation, yet are still present and easily observable today, and in particular yield ages that are very precise. 
It has been well established that the Galactic bulge Globular Clusters (BGCs) form an independent system from that of the halo \citep{Minniti1995} and were most likely formed in situ as opposed to the ex situ halo GCs (e.g. Massari et al. 2019), and are therefore very valuable for investigating the oldest native Milky Way relics and its formation.
In 2016, the VISTA Variables in the Via Lactea Survey (VVV survey) contained more than 55 GCs in the Bulge area of the Milky Way (MW) as \cite{bica_ortolani_barbuy_2016} reported, and a number of new candidates have recently been reported by Bica et al. (2024). Many of these are relatively unknown and poorly studied due to visibility limitations of this region, caused by heavy dust obscuration. 

{  The Apache Point Observatory Galactic Evolution Experiment II survey \citep[APOGEE-2;][]{Majewski_2017AJ....154...94M}, part of the Sloan Digital Sky Survey-IV \citep{Blanton2017}, was developed to provide precise radial velocities (RV $<$1 km s$^{-1}$) and detailed chemical abundances for an unprecedentedly large sample of giant stars, aiming to unveil the dynamical structure and chemical history of all  MW components. 
The bulge Cluster APOgee Survey (CAPOS) was 
granted time as an External Program to APOGEE-2, focused on studying the formation and evolution of the Galactic Bulge, using a large sample of BGCs located in the inner $\pm$10$^\circ$ $\times$ $\pm$10$^\circ$ around the Galactic center \citep{Geisler_2021_2021A&A...652A.157G}. CAPOS obtained high-resolution ($R\sim$22,500) spectra  in the Near Infrared (\textit{H}-band), thus able to penetrate dust much more than previous observations limited to the optical. CAPOS observations were carried out on the Ir\'en\'ee du Pont 2.5m telescope \citep[][]{Bowen1973} at Las Campanas Observatory (APOGEE-2S). The APOGEE spectra deliver detailed abundance and kinematic information of individual giant stars with an internal accuracy in chemical abundances better than  $\sim$0.05 dex and radial velocities with error $<$ 0.1 km s$^{-1}$
for well-observed (S/N$\geq 70$) stars.  Each spectra provides information on more than 25 chemical species, including: C, C I, N, O, Na, Mg, Al, Si, P, S, K, Ca, Ti, Ti II, V, Cr, Mn, Fe, Co, Ni, Cu, Ge, Rb, Ce II, Nd II, and Yb II \citep{Shetrone2015, Hasselquist2016,  Cunha2017, Smith2021}.}

The { CAPOS GC targets} were taken from \citep{Harris_2010arXiv1012.3224H}. In selecting the CAPOS sample, a number of criteria were considered, including prioritizing the most metal-poor BGCs, those that were poorly studied, new GC candidates \citep{Minniti_a_2017RNAAS...1...16M,Minniti_b_2017ApJ...849L..24M,Barba_2019ApJ...870L..24B,Palma_2019MNRAS.487.3140P} and maximizing the number of observable clusters per field. A total of 18 BGCs were observed in CAPOS. The sample includes GCs that were classified as BGCs taking into account only their Galactic location \citep{bica2016} but not kinematics, since the observations were generally obtained before the release of Gaia DR2, which provided orbital characteristics of GCs for the first time
\citep{Massari2019}. CAPOS is particularly focused on the chemical abundances of [Fe/H], [$\alpha$/Fe] and [C,N,O/Fe] for BGCs that are key 
to supplement deep photometry to derive ages from isochrones, since ages strongly depend on these values. All these parameters will contribute to our knowledge of Galactic formation and chemical evolution.

Initial results based on ASPCAP, APOGEE's internal piepline, abundances for a subsample of the earliest CAPOS clusters observed were presented in \cite{Geisler_2021_2021A&A...652A.157G}. Several CAPOS clusters were subsequently studied in more detail, including FSR~1758 \citep{Romero_2021A&A...652A.158R}, where we reported a new paradigm for this GC, { NGC~6380 \citep[Tonantzintla~1 ;][]{Fernandez-Trincado2021a} and Tonantzintla~2 \citep{Fernandez-Trincado2022b} where evidence for a correlation between light- and slow neutron-capture process (Ce II) elements was identified}, and NGC~6558  \citep{2023NGC6558} {  where a detailed chemical abundance analysis was reported for O$^{16}$H, C$^{12}$O$^{16}$ and C$^{12}$N$^{14}$ molecules}. In addition, the presence of the multiple-population (MP) phenomenon in all of these objects was investigated.
{ Elemental abundances in GCs from APOGEE-2 data have been analyzed in a series of papers by \citet{Meszaros2015}, \citet{Schiavon2017},  \cite{Masseron2019}, \citet{Fernandez-Trincado2019, Fernandez-Trincado2020a, Fernandez-Trincado2020b, Fernandez-Trincado2021a, Fernandez-Trincado2021b, Fernandez-Trincado2021c, Fernandez-Trincado2021d, Fernandez-Trincado2022a, Fernandez-Trincado2022b}, \citet{Meszaros_2020}, \citet{Geisler_2021_2021A&A...652A.157G}, \cite{Meszaros_2021MNRAS.505.1645M}, and \citet{Schiavon2024} showing evidence of chemical inhomogeneities and MPs in all clusters investigated.}

{ The MP phenomenon has been found across the entire parameter space covered by GCs and their chemical inhomogeneities identified and mapped. MPs have been also widely examined by \cite{Carreta_2009A&A...505..117C} and \citet[][and references therein]{Bastian2018},} where the authors find a significant spread in Na and O, indeed an anti-correlation in all their targets.  These abundance spreads are presumably due to the early evolution of each cluster, formed initially by a first population of stars that has the same chemical composition as field stars at the same metallicity. The second (likely subsequent) population of stars (Na-richer and O-poorer) is formed from gas polluted by ejecta of evolved stars of the first population (Na-poorer and O-richer). This is the so-called MP phenomenon. This spectroscopic evidence has been interpreted as the signature of material processed during H-burning at high temperatures by proton-capture reactions (such as the Ne–Na and Mg–Al cycles). Several sources of polluters have been proposed: intermediate-mass asymptotic giant branch stars  \citep{D'ANTONA_2016MNRAS.458.2122D}, fast-rotating massive stars \citep{DECRESSIN_2007A&A...464.1029D}, massive binaries \citep{deMink_2009A&A...507L...1D} and early disk accretion \citep{BASTIAN_2013MNRAS.436.2398B}.
However, none of the proposed scenarios to date can account for all observational constraints, as explained by \citet{Renzini_2015MNRAS.454.4197R}, demanding both additional theories as well as observations. { BGCs are especially important in this regard as they open up the realm of higher metallicity, which is not populated by halo GCs.}

\begin{figure}[thpb]
\centering
\includegraphics[scale=0.4]{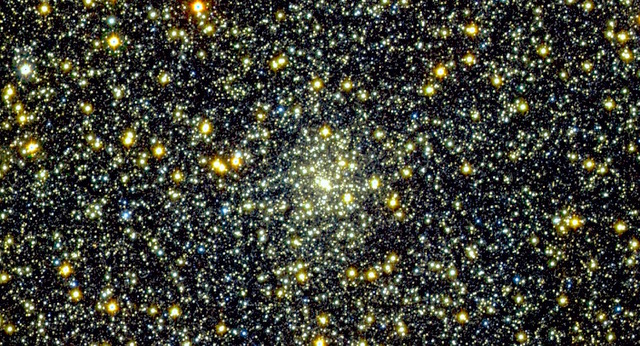}
   \caption{Multi-band (JHKs - combined colour) infrared view of HP~1. Image from the Vista Variables in the Via Lactea (VVV) survey. 
   }
\label{Vista_Hp1_VVV}
\end{figure}

{ Among the CAPOS BGCs, HP~1 is one of the more intriguing and possibly oldest clusters, which has been examined photometrically} by \cite{Kerber2019}, and constitutes a fossil relic of the Galactic bulge. Figure \ref{Vista_Hp1_VVV} displays a VVV image of HP~1  from \citet{Saito_2012_hp1_vvv_2012A&A...537A.107S}, clearly showing its globular nature.

\begin{figure*}[h]
	\centering
	\includegraphics[width=1.0\textwidth]{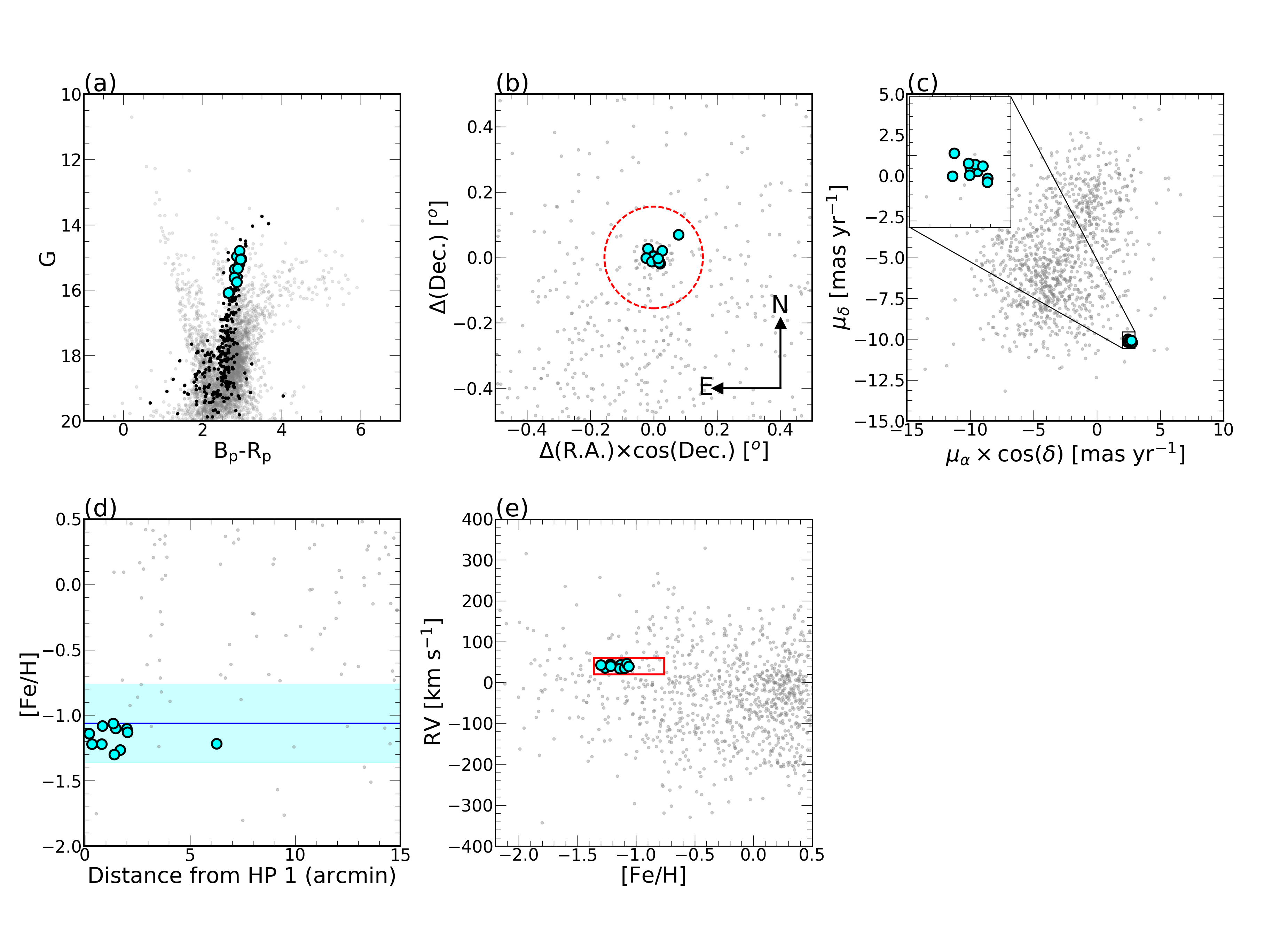}
	\caption{ Main properties of HP~1 stars. Panel (a): Color-magnitude diagram in the Gaia bands for cluster stars with a membership probability larger than 90\% (black dots) and $<$90\% (grey dots) taken from \citet[][]{Baumgardt_2021}. The eleven stars analyzed in this work are marked with cyan symbols  in all the panels. The APOGEE-2 footprint stars toward HP~1 are marked with grey dots in panel (b), (c), (d) and (e). Panel (b): Sky position of stars centered on HP~1; the overlaid large red dashed circle refers to the cluster tidal radius (${\rm r_{t}}$). Panel (c): PM distribution for stars toward the field of HP~1, with the inner zoomed window highlighting the distribution of our sample within a 0.5 mas yr$^{-1}$ radius around the nominal PM of the cluster. Panel (d): [Fe/H] versus the projected angular cluster distance, with the blue line showing the assumed initial mean cluster  metallicity  and the cyan shadow region indicating $\pm0.3$ dex from the initial metallicity. Panel (e): Radial velocity versus [Fe/H] of our member stars compared to APOGEE-2 field stars. The red box limited by $\pm$0.30 dex and $\pm$20 km s$^{-1}$, centered on [Fe/H] $= -1.06$ and RV $= +39.76$ km s$^{-1}$, encloses our potential cluster members (see Sect. \ref{Methods}). The metallicity values shown in panels (d) and (e) have been taken from the \texttt{ASPCAP} pipeline.  }
	\label{fig2}
\end{figure*}

HP~1 is a bona fide GC with a reasonably low metallicity for a BGC and also possesses a BHB,  with an estimated age of 12.8 $_{-0.8}^{+0.9}$ Gyr \citep{Kerber2019}, a mass of 1.2 $\times10^5$ M$\odot$, a distance from the sun of 7000 $\pm$ 140 pc, a Galactocentric distance of 1260 $\pm$ 130 pc, and a tidal radius of 19 pc, as reported by \cite{Baumgardt_2021}. It is one of the most centrally located GCs in the Milky Way,  projected at 3.3$^\circ$ from the Galactic center, with a reported metallicity of [Fe/H] $\sim$ $-$1.0 (Harris 2010) \citep{Barbuy2016A&A...591A..53B}. In the latter study, the authors obtained abundances of the elements C, N, O, Na, Mg, Al, Si, Ti and Fe. $\alpha$ elements are enhanced, as expected, with mean [O,Mg,Si/Fe] values of $+$0.40, $+$0.36 and $\lesssim$ $+$0.35, respectively,
while Ti shows a mean abundance of [Ti/Fe] $\lesssim$ $+$0.16.

CAPOS data provides much better and completely self -consistent spectroscopic metallicity than any currently available for HP~1, as well as detailed abundances for many elements with a wide variety of nucleosynthetic origins, including the { $\alpha$- (O, Mg, Si, S, Ca, Ti), Fe-peak (Fe, Ni), light- (C, N),  odd-Z (Al), and \textit{s}-process elements (Ce).
Geisler et al. (2021) derived a mean metallicity of -1.20$\pm 0.10$ from ASPCAP DR16 abundances.}

In this study, we  make use of the \texttt{BACCHUS} (the Brussels Automatic Code for Characterizing High accUracy Spectra) package \citep{Masseron2016} in order to characterize the chemical abundances in HP~1. We made a careful line-by-line selection in each spectra. For each line of each element \texttt{BACCHUS} synthesizes 
spectra with different abundances and returns the best fit value.

{ We organize this paper in the following way:} in Sect. \ref{Methods} we describe the methods and data. In Sect. \ref{stellar_param} we review HP~1 and also describe the stellar parameters. Sect. 4 presents the abundance determinations.
The results are described in Sect. \ref{Results}, and finally, we present our summary and concluding remarks in Sect. \ref{summary}.

\begin{table}
	\caption[]{Global properties of HP~1 taken from \citet{Baumgardt_2021} and \citet{Harris_1996AJ....112.1487H} -- 2010 edition. }
	\label{table1}
	$$ 
	\begin{array}{p{0.5\linewidth}l}
	\hline
	\noalign{\smallskip}
	\text{Positional Parameters}& \text{Derived value} \\
	\noalign{\smallskip}
	\hline
	\noalign{\smallskip}
	\textit{$\alpha_{\rm J2000}$} \text{(hh:mm:ss)} & \text{17:31:05.2}  \\ 
	\textit{$\delta_{\rm J2000}$} \text{(deg:min:seg)} & \text{$-$29:58:54}  \\ 
	\textit{\text{{R}\textsubscript{$\odot$}}} & \text{7.0} \pm \text{0.14}\text{ kpc}\\ 
	\textit{R\textsubscript{{GC}}} & 1.26 \pm \text{0.13}\text{ kpc}\\
	\textit{L} \text{(galactic longitude)}& \text{357.425} \text{ deg} \\
	\textit{B} \text{(galactic latitude)} &\text{2.115} \text{ deg} \\
	\noalign{\smallskip}
	\hline
	\noalign{\smallskip}
	Structural Parameters\\
	\noalign{\smallskip}
	\hline
	\noalign{\smallskip}
	\textit{M} & 1.24 \pm0.17 \times 10^{5} \text{ M} \textsubscript{$\odot$} \\
	\textit{r\textsubscript{m}} & \text{3.74 pc} 
	\\
	\textit{r\textsubscript{c}} \text{(core radius)} & \text{1.26 pc} \\ 
	\textit{R\textsubscript{V}} \text{(radial velocity)} & 39.76 \pm \text{1.22 km s}^{-1} \\ 
	\textit{r\textsubscript{t}} \text{(tidal radius)}& \text{19 pc}\\
	\noalign{\smallskip}
	\noalign{\smallskip}
	\hline
	\noalign{\smallskip}
	Photometric Parameters\\
	\noalign{\smallskip}
	\hline
	\noalign{\smallskip}
	\textit{E(B-V)} & \mathrm{1.12}\\
	\textit{(m-M)\textsubscript{V}} & 18.05 \\
	\noalign{\smallskip}
	\hline
	\noalign{\smallskip}
	\end{array}
	$$ 
\end{table}

\begin{table*}
	\begin{small}
		\begin{center}
			\setlength{\tabcolsep}{0.8mm}  
			\caption{Photometry, kinematics and astrometric properties of the 14 likely HP~1 members.}
			\begin{tabular}{cccccccccccccc}
				\hline
					\noalign{\smallskip}
				APOGEE\_ID  & FIELD & S/N & $\alpha$ & 
				$\delta$ & J & H &  K$_{\rm s}$ & BP & RP & G & RV$_{\rm H}$ & $\mu_{\alpha}\cos(\delta)$ & $\mu_{\delta}$\\
				& & pixel$^{-1}$  & hh:mm:ss & dd:mm:ss & & & & & & & km s$^{-1}$ & mas yr$^{-1}$ & mas yr$^{-1}$\\
					\noalign{\smallskip}
				\hline
					\noalign{\smallskip}
				2M17311037$-$2957146 & 357$+$02$-$C & 129.22 & 17:31:10.38 & $-$29:57:14.66 & 12.19 & 11.16 & 10.86 & 17.23 & 14.37 &  15.60 & 36.93 & 2.426 & $-$10.158 \\
				2M17305949$-$3000019 & 358$+$02$-$O & 156.53 & 17:30:59.49 & $-$30:00:01.91 & 11.55 & 10.56 & 10.29 & 16.59 & 13.72 & 14.97 & 37.00 & 2.626 & $-$10.070 \\
				2M17305774$-$2957404 & 357$+$02$-$C & 100.86 & 17:30:57.75 & $-$29:57:40.46 & 11.92 & 10.90 & 10.60 & 16.94 & 14.12 & 15.36 & 43.83 & 2.441& $-$9.984  \\
				2M17310585$-$2958354 & 357$+$02$-$C & 122.10 & 17:31:05.85 & $-$29:58:35.48 & 12.21 & 11.27 & 10.97 & 17.12 & 14.32 & 15.61 & 44.38 & 2.771	& $-$10.174	\\
				2M17310541$-$2958416 & 357$+$02$-$C & 131.93 & 17:31:05.42 & $-$29:58:41.67 & 11.22 & 10.16 & 9.85 & 16.44 & 13.51 & 14.79 & 34.63 & 2.673	& $-$10.122 \\
				2M17311196$-$2958587 & 357$+$02$-$C & 115.60 & 17:31:11.97 & $-$29:58:58.70 & 11.68 & 10.59 & 10.29 & 16.85 & 13.93 & 15.20 & 34.99 & 2.600 & $-$10.098 \\
				2M17310703$-$2959376 & 357$+$02$-$C & 133.73 & 17:31:07.04 & $-$29:59:37.62 & 11.47 & 10.36 & 10.05 & 16.77 & 13.79 & 15.04 & 46.19 & 2.767 & $-$10.204	\\
				2M17310020$-$2959415 & 357$+$02$-$C & 89.82 & 17:31:00.21 & $-$29:59:41.58 & 11.85 & 10.78 & 10.52 & 16.99 &  14.10 & 15.33 & 39.35 & 2.649 & $-$10.067\\
				2M17310045$-$2959502 & 357$+$02$-$C & 115.57 & 17:31:00.45 & $-$29:59:50.30 & 11.40 & 10.42 & 10.20 & 16.78 & 13.81 & 15.06 & 42.97 & 2.582 & $-$10.060\\
				2M17304363$-$2954441 & 358$+$02$-$O &  98.51 & 17:30:43.64 & $-$29:54:44.19 & 12.26 & 11.31 & 11.01 & 17.36 & 14.49 & 15.75 & 40.65 & 2.724 & $-$10.080 \\ 
					\noalign{\smallskip}
				\hline 
					\noalign{\smallskip}
					2M17310160$-$2959048 & 357$+$02$-$C & 74.43 & 17:31:01.60 & $-$29:59:04.85 & 12.81 & 11.91 & 11.68 & 17.49 & 14.83 & 16.08 & 41.82 & 2.592 & $-$10.151		\\
					\noalign{\smallskip}
				\hline
					\noalign{\smallskip}
				2M17310872$-$3000217 & 357$+$02$-$C & 64.22 & 17:31:08.73 & $-$30:00:21.74 & 11.81 & 11.13 & 10.76 & 16.27 & 13.62 & 14.84 & 50.87 & 2.645 & $-$9.991 	\\
				AP17310188$-$2959412 & 358$+$02$-$O & 15.01 & 17:31:01.88 & $-$29:59:41.20 & ... & 14.17 & ... & 18.79 & 16.37 & 17.77 & 31.16 & 2.981 & $-$9.971 \\
				AP17310775$-$2958508 & 358$+$02$-$O & 11.98 & 17:31:07.75 & $-$29:58:50.80 & ... & 14.36 & ... & 19.16 &  16.65 & 17.93 & 32.63 & 2.167 & $-$10.317 \\
					\noalign{\smallskip}
				\hline
			\end{tabular}  \label{table2}
		\end{center}
	\end{small}
\end{table*}   

\section{Data and sample}\label{Methods}

{ Our HP~1 CAPOS spectra are part of the final 17th data release of SDSS-IV \citep{Abdurro2022}, which includes all APOGEE-2 data. Spectra were reduced as described in \citet{Nidever2015}, and analyzed using the APOGEE Stellar Parameters and Chemical Abundance Pipeline \citep[\texttt{ASPCAP};][]{Garcia2016}, and the libraries of synthetic spectra described in \citet{Zamora2015}. The customised \textit{H}-band line lists are fully described in \citet{Shetrone2015, Hasselquist2016, Cunha2017} and \citet{Smith2021}.}
{ The APOGEE-2S plug-plates containing HP~1 members were centered on ($l$,$b$) $\sim$ (357$^{\circ}$, $+2.0^{\circ}$) and (358$^{\circ}$, $+2.0^{\circ}$ ), which correspond to the APOGEE-2 programs of \texttt{geisler\_18a} (plate name 357$^{\circ}+2.0^{\circ}$C), \texttt{schlaufman\_17a} and \texttt{kollmeier\_17a} (plate name 358$^{\circ}+2.0^{\circ}$O).}

{ We adopt the following criteria to select our final HP~1 members, which are based on those described in \cite{Geisler_2021_2021A&A...652A.157G}, \cite{Romero_2021A&A...652A.158R}, \cite{Fernandez-Trincado2022b}: \textit{i}) stars located within the tidal radius of 19 pc (30 arcmin) from the cluster center; \textit{ii}) stars with proper motions (PMs) within a radius of 0.5 mas yr$^{-1}$ from the mean cluster value; \textit{iii}) stars with radial velocity within 20 km s$^{-1}$ from the cluster mean radial velocity, 39.76 km s$^{-1}$ taken from \citep{Baumgardt_2021}; and \textit{iv}) stars differing in [Fe/H] from the assumed mean value \citep[$-1.06$;][]{Barbuy2016A&A...591A..53B} by  $<$0.3 dex. This [Fe/H] value was used as our initial metallicity, which was re-calculated in this work (see Sect. \ref{Results}). Fourteen stars satisfied all of these criteria and were  considered our final members. 
	
Table \ref{table1} lists the main physical parameters  adopted for HP~1, while Table \ref{table2} lists  the 14 members. However, we chose to analyze only the 11 stars which have a  signal-to-noise ratio (S/N)  $>$70 pixel$^{-1}$. This limit for S/N has been suggested by the APOGEE team 
(e.g. Meszaros et al. 2020) and 
\cite{Fernandez-Trincado2021a, Fernandez-Trincado2022b} in order to guarantee reliable abundance determinations. Stars which do not meet this requirement are listed in the last three rows in Table \ref{table2}. We also removed one additional star, 2M17310160$-$2959048,  from our abundance analysis, as a visual inspection of its spectrum reveals that the Fe I lines are very weak, noisy, and with unusual line profiles limiting our ability to produce reliable abundances. This star is listed in the eleventh row in Table \ref{table2}. Note that the \texttt{ASPCAP} pipeline provides some estimates of abundances for this star, which we suspect are not reliable. As an initial guess, we adopted the [Fe/H] abundance ratios from \texttt{ASPCAP}, which are also shown in Figure \ref{fig2} and Table \ref{table3}. 
}

{ Figure \ref{fig2} summarizes the photometric, kinematic, astrometric, and metallicity  properties of the 11 final HP~1 members
analyzed. }

\section{Stellar parameters} \label{stellar_param}

\begin{figure}[h]
	\centering
	\includegraphics[scale=0.45]{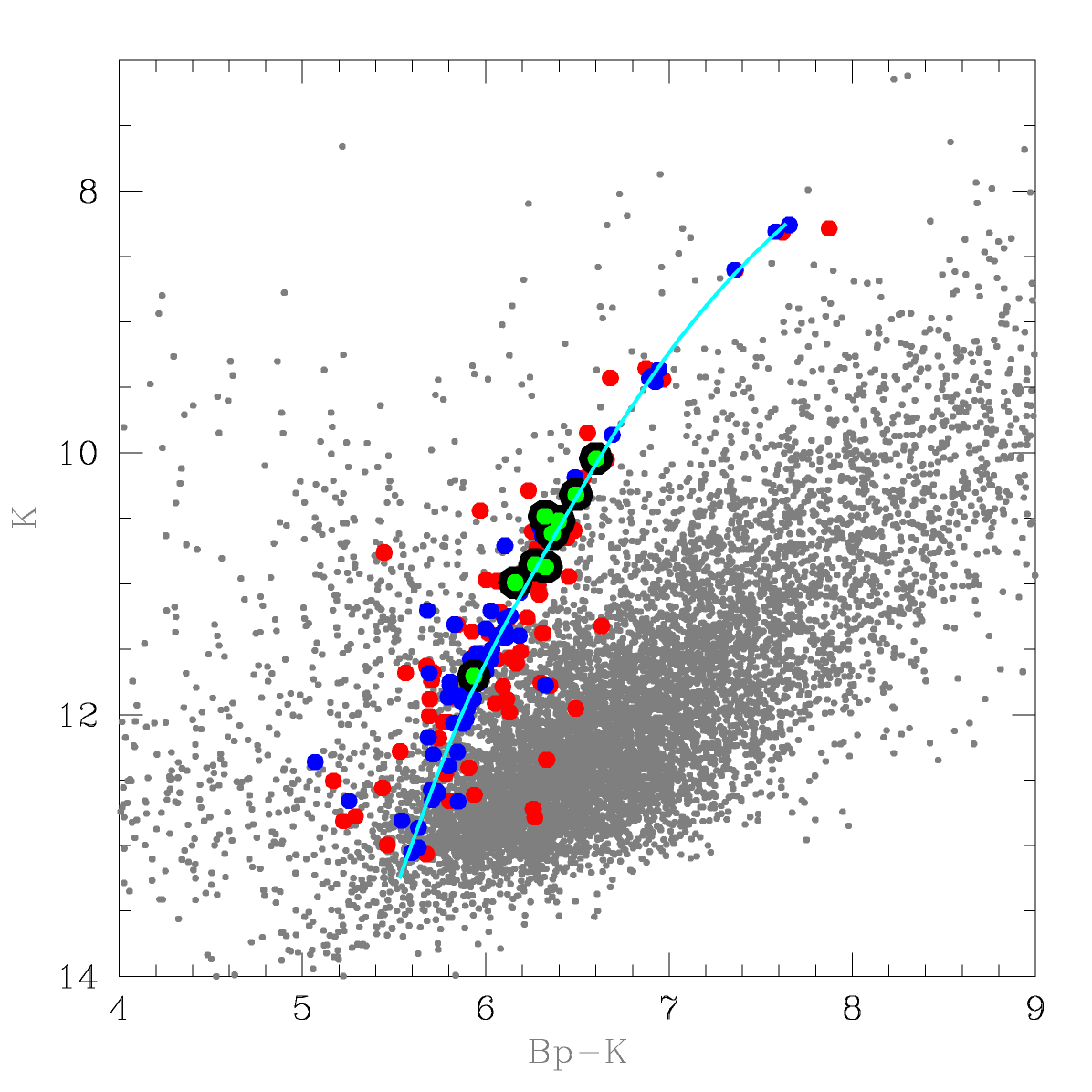}
	\caption{\bf K vs. B$_p$ $-$ K CMD of HP~1. The cyan line show a portion along the RGB sequence of a 13.0 Gyr isochrone with [M/H]= $-$0.90 shifted using E(B$-$V)$=$1.01, and (m-M)$_0$ $=$ 14.10. Gray points represents non member stars (according to proper motions). Red circles represent member stars while blue circles are member stars corrected for differential
reddening. Finally green circles represent our targets.}
	\label{CMD_parameters}
\end{figure}

The atmospheric parameters of our targets were obtained through an iterative procedure based on the GAIA (G, B$_{\rm p}$, and R$_{\rm p}$) and 2MASS (J, H, and $K_S$) photometry of HP~1. During this procedure, the effective temperature (T$_{\text{eff}}$) and gravity (log g) of each red giant branch (RGB) star (including our targets) were obtained, and at the same time, colour magnitude diagrams (CMDs) were corrected for differential reddening. First, we fit a \texttt{PARSEC} isochrone \citep{2012MNRAS.427..127B} to the RGB, assuming an age of 13.0 Gyrs as shown in Figure \ref{CMD_parameters}. We accounted for reddening by applying the \citet{Cardelli1989} reddening law to the isochrone. The visual absorption A$_V$, the R$_V$ parameter, the intrinsic distance modulus (m-M)$_0$ and the global metallicity [M/H] were determined by simultaneously fitting the RGB, RC and RGB-tip in the K vs.  B$_{\rm p}$ $-$ K, G vs.  B$_{\rm p}$ $-$ R$_{\rm p}$, B$_{\rm P}$ vs. B$_{\rm p}$ $-$ R$_{\rm p}$ and K vs. J $-$ K CMDs. The T$_{\text{eff}}$ and $\log$ \textit{g} of each RGB star were then determined as those corresponding to the point on the isochrone where the $K_s$ magnitude matches that of the star (we avoided using RC stars for this step). We used the K$_s$ magnitude because it is the least affected by reddening and, consequently, by differential reddening. Having the temperature we obtained the intrinsic B$_{\rm p}$ $-$ K$_s$ color of each star from the color-temperature relation of the RGB part of the isochrone and, by subtracting the mean reddening obtained from the isochrone fitting, also the differential reddening at the position of each star. Finally, for each star we selected the 4 closest neighbours (5 stars in total) and corrected its G, B$_{\rm p}$, R$_{\rm p}$, J, H, K magnitudes using the mean differential reddening of the 5 stars. We used the B$_{\rm p}$ $-$ K color because it is the most sensitive to any reddening variation. This procedure was iterated until any improvement of the CMDs was negligible.

We obtained A$_V$ = 3.8, R$_V$ = 2.4, (m$-$M)$_0$ = 14.10 and   [M/H] = $-$ 0.9 [dex], a value higher than the iron content of the cluster ([Fe/H] = $-$1.1). This is not surprising since globular clusters are usually $\alpha$-enhanced.

The microturbulence ($\xi_t$) was determined  using the empirical equation given in \cite{Dutra2016}:
\begin{equation}
\xi_t = 0.998+3.16\times 10^{-4}X-0.253Y-2.86\times 10^{-4}XY+0.165Y^2 
\label{eq1}
\end{equation}
where X = \textit{T}\textsubscript{eff} $-$ 5500 [K] and Y $= \log (g) - 4.0$.  
As \cite{Dutra2016} indicate, this equation is consistent with the microturbulence values computed from 3D models. We refer the reader to these papers for detailed information.

Table \ref{table3} lists our derived atmospheric parameters (T$_{\rm eff}$, log g,
and $\xi_t $) for the 11 HP~1 stars, which are compared to those derived from the \texttt{ASPCAP} pipeline. 

\section{Abundance determinations}\label{abundances}

We made use of the Brussels Automatic Stellar Parameter (\texttt{BACCHUS}) code \citep{Masseron2016} to manually analyze each star of our sample, in order to examine the reliability of each atomic and molecular line present in each spectrum, and derive metallicity (from Fe I lines), broadening parameters, and  chemical abundances for 12 chemical species, listed in Table \ref{table3}. 

{ The \texttt{BACCHUS} code relies on the radiative transfer code \texttt{Turbospectrum} \citep{Plez2012} and the $\alpha$-rich ([$\alpha$/Fe]$=+0.4$) \texttt{MARCS} model atmosphere grid \citep{Gustafsson2008}, and the abundances are computed adopting a line-by-line approach under the assumption of local thermodynamic equilibrium (LTE). For each element and each line, the abundance determination proceeds as in our previous CAPOS  BACCHUS papers \citep{Fernandez-Trincado2021a}. In summary, the steps are: (\textit{i}) a spectrum synthesis, using the full set of (atomic and molecular) lines to find the local continuum level via a linear fit; (\textit{ii}) cosmic and telluric line rejections are performed; (\textit{iii}) the local signal-to-noise ratio (S/N) per element is estimated; (\textit{iv}) a series of flux points contributing to a given absorption line are automatically selected; and (\textit{v}) abundances are then derived by comparing the observed spectrum with a set of convolved synthetic spectra characterised by different abundances.}

{ Four different abundance determinations are used: (\textit{i}) line-profile fitting; (\textit{ii}) core line-intensity comparison; (\textit{iii}) equivalent-width comparison; and (\textit{iv}) global goodness-of-fit estimate. Each diagnostic yields validation flags. Based on these flags, a decision tree then rejects or accepts the line, keeping the best-fit abundance. We adopted the $\chi^{2}$ goodness-of-fit diagnostic as the final abundance, because it is considered to be the most robust. However, we stored the information from the other diagnostics, including the standard deviation between all four methods. The line list used in this work is the latest internal DR17 atomic/molecular linelist (\texttt{linelist.20170418}), including the $s$-process elements (Ce II, Nd II, and Yb II)--\citep{Hasselquist2016, Cunha2017}. }

{ In particular, a mix of heavily CN-cycled and $\alpha$-rich \texttt{MARCS} models were used, as well as the same molecular lines adopted by APOGEE-2 \citep{Smith2013}, in order to determine the C, N, and O abundances.  In addition, we have adopted the C, N, and O abundances that satisfy the fitting of all molecular lines consistently; that is to say, we first derive $^{16}$O abundances from $^{16}$OH lines, then derive $^{12}$C from $^{12}$C$^{16}$O lines, and $^{14}$N from $^{12}$C$^{14}$N lines; the C--N--O abundances were derived iteratively to minimize the $^{16}$OH, $^{12}$C$^{16}$O, and $^{12}$C$^{14}$N dependences \citep{Smith2013}.}

{ The adoption of a purely photometric temperature scale enables us to be somewhat independent of the \texttt{ASPCAP}/APOGEE-2 pipeline, which provides important comparison data for validation. The final results presented in this paper are based on computations done with the \texttt{BACCHUS} code using the listed photometric atmospheric parameters in Table  \ref{table3}. The same table also lists our resulting elemental abundances for the targets analyzed in this work. }

\begin{figure}
	\centering
	\includegraphics[scale=0.18]{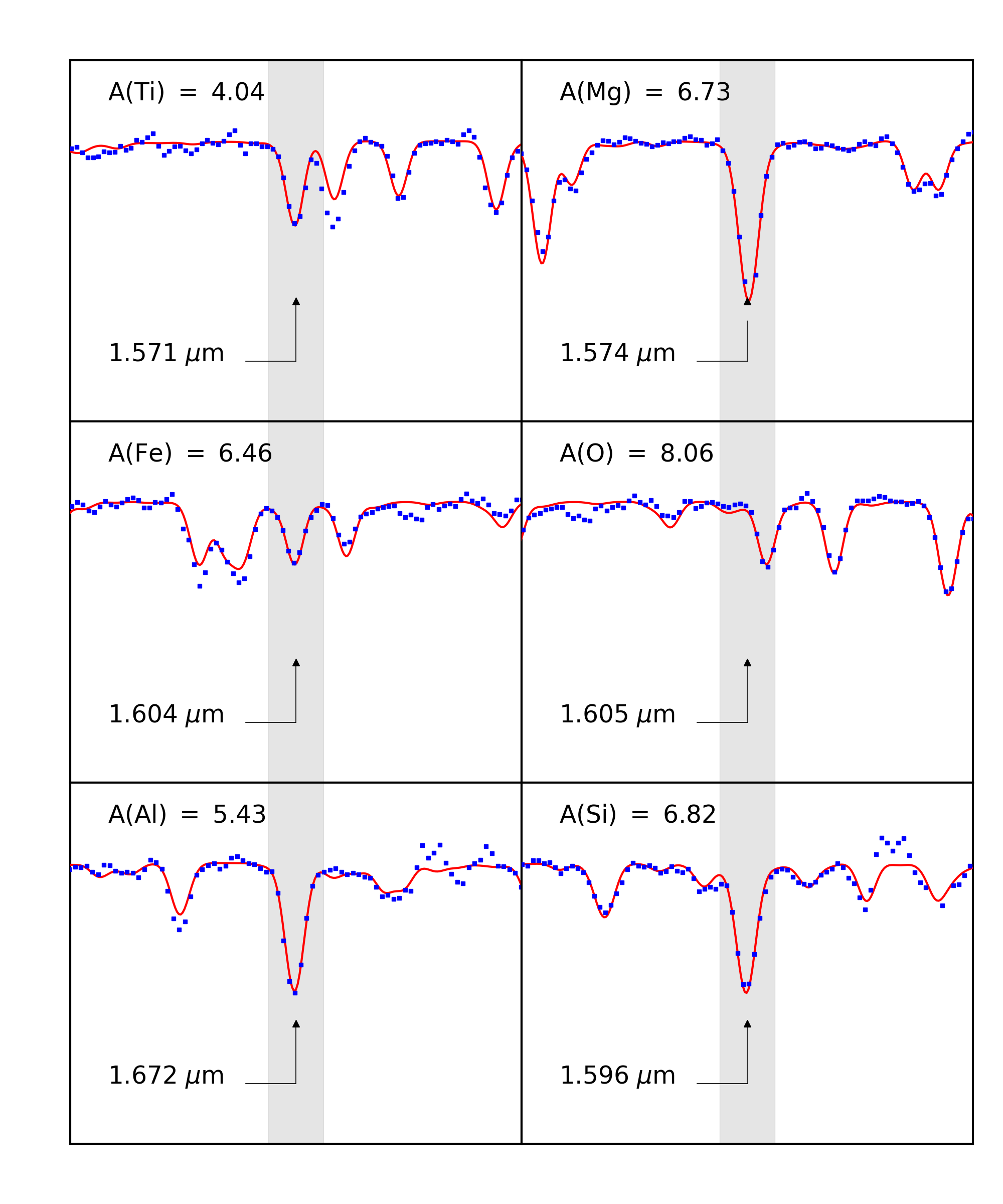}
	\caption{ Comparison of synthetic spectra (red lines) to the observed spectra (blue dots) for a HP~1 star, 2M17310541$-$2958416. Each panel show the best-determined (grey shadow region) [Ti/Fe], [Mg/Fe], [Fe/H], [O/Fe], [Al/Fe] and [Si/Fe] abundance ratios listed in Table \ref{table3}. An arbitrary normalized flux is plotted in the vertical axis, while the air wavelength ($\mu$m) is plotted in the horizontal axis.}
	\label{spectra}
\end{figure}

{ Figure \ref{spectra} shows an example of our best \texttt{BACCHUS} spectral synthesis on clean selected features for star 2M17310541$-$2958416. }

\begin{table*}
	\begin{small}
		\begin{center}
			\setlength{\tabcolsep}{0.7mm}  
			\caption{\bf \texttt{BACCHUS} and \texttt{ASPCAP} elemental abundances of 10 members in HP~1.}
			\begin{tabular}{cccccccccccccccc}
				\hline
					\noalign{\smallskip}
	APOGEE\_ID   & 	T$_{\rm eff}$ & $\log$ \textit{g} & $\xi_t$ & [C/Fe] & [N/Fe] &  [O/Fe] & [Mg/Fe] & [Al/Fe] & [Si/Fe] & [S/Fe]$^*$ & [Ca/Fe] & [Ti/Fe] &  [Fe/H]  & [Ni/Fe] & [Ce/Fe]$^\dag$\\ 
			& K & c.g.s & km s$^{-1}$ & &&&&&&&&&&&\\
				\noalign{\smallskip}
            \hline
            	\noalign{\smallskip}
\texttt{BACCHUS} results & & &  &   &   &    &      &  &    &     &  &    &   &   & \\
	\noalign{\smallskip}
			\hline
				\noalign{\smallskip}
			2M17311037$-$2957146 & 4378 & 1.47 & 1.53 & $-$0.33  & $+$1.03 & $+$0.34 & $+$0.32   & $+$0.57   & $+$0.38   & ...     & $+$0.27   &  $+$0.21   & $-$1.04   & $+$0.05   & ... \\
			2M17305949$-$3000019 & 4250 & 1.24 & 1.57 & $-$0.26  & $+$0.89 & $+$0.30 & $+$0.26$^*$ & $+$0.23   & $+$0.48   & ...     & $+$0.09$^*$ &     ...    & $-$1.21   & $-$0.01$^*$ & ...  \\
			2M17305774$-$2957404 & 4321 & 1.37 & 1.55 & $-$0.36  & $+$1.32 & $+$0.36 & $+$0.37   & $+$0.52   & $+$0.36   & $+$0.59 & $+$0.35   &  $+$0.21   & $-$1.16   & $-$0.04   & $+$0.25 \\
			2M17310585$-$2958354 & 4413 & 1.53 & 1.52 & $-$0.21  &    ...  & $+$0.25 & $+$0.20   & $+$0.07   & $+$0.44   & ...     & $+$0.02$^*$ &     ...    & $-$1.21   & $+$0.05   & ...  \\
			2M17310541$-$2958416 & 4139 & 1.04 & 1.61 & $-$0.11  & $+$0.38 & $+$0.56 & $+$0.35   & $+$0.15   & $+$0.39   & ...     & $+$0.30   &  $+$0.23   & $-$1.12   &   ...     & $+$0.17 \\
			2M17311196$-$2958587 & 4241 & 1.22 & 1.58 & $-$0.08  & $+$1.01 & $+$0.31 & $+$0.33   & $+$0.61   & $+$0.39   & ...     &    ...    &  $+$0.23$^*$ & $-$1.23   & $+$0.15   & ...  \\
			2M17310703$-$2959376 & 4183 & 1.12 & 1.59 & $-$0.61  & $+$1.25 & $+$0.07 & $+$0.20   & $+$0.88   & $+$0.40   & ...     & $+$0.32   &  $+$0.20   & $-$1.13   & $+$0.01   & $+$0.16$^*$ \\
			2M17310020$-$2959415 & 4299 & 1.32 & 1.56 & $-$0.52  & $+$1.31 & $+$0.06 & $+$0.33$^*$ & $+$0.91   & $+$0.55$^*$ & ...     & $+$0.38$^*$ &  $+$0.12   & $-$1.02   & ...       & ...  \\
			2M17310045$-$2959502 & 4218 & 1.18 & 1.58 & $+$0.03  & $+$0.37 & $+$0.46 & $+$0.22$^*$ & $+$0.23$^*$ & $+$0.50$^*$ & ...     &    ...    &  ...       & $-$1.08   & ...       & ... \\
			2M17304363$-$2954441 & 4414 & 1.53 & 1.52 & $-$0.25  &    ...  & $+$0.30 & $+$0.30   & $+$0.43$^*$ & $+$0.63   & $+$0.59 &    ...    &  ...       & $-$1.27   & $+$0.03$^*$ & ...  \\  
				\noalign{\smallskip}
			\hline
				\noalign{\smallskip}
			Median & ... & ... & ... &    $-$0.26&    $+$1.02  &   $+$0.30 &     $+$0.31  &     $+$0.47 &    $+$0.42  &  $+$0.59  &   $+$0.30 &    $+$0.21&   $-$1.15   &  $+$0.03  &  $+$0.17  \\
			\bf{Mean} & ... & ... & ... & $-$0.27 &   $+$0.94  &   $+$0.30 &     $+$0.29  &     $+$0.46 &    $+$0.45  &  $+$0.59  &   $+$0.25 &    $+$0.20 &  $-$1.15   &  $+$0.03  &  $+$0.19 \\
			1$\sigma$ & ... & ... & ... &   0.18  &  0.43  &   0.13 &     0.07  &     0.29 &    0.07  &  ...      &   0.13 &    0.02 &  0.08   &  0.03  &  0.03              \\      
			std & ... & ... & ... &   0.19  &  0.36  &   0.15 &     0.06  &     0.28 &    0.08  &  ...      &   0.13 &    0.04 &  0.08   &  0.06  &  0.04              \\      			
			spread & ... & ... & ... &    0.64 &  0.95  &   0.50 &  0.17     &      0.84 &     0.27 &  ...      &    0.36 &     0.11 &   0.25 & 0.11 &   0.09             \\     
			\hline 
							\noalign{\smallskip} 
						$\sigma_{\rm total}$  (2M17305949$-$3000019) & ... & ... & ... &    0.10 & 0.11 &    0.12 &      0.09  &      0.10 &     0.10 &  ...      &    0.08 &     0.13 &   0.22 &  0.10 &   0.11             \\      
				\noalign{\smallskip} 
			\hline
				\noalign{\smallskip}
			\texttt{ASPCAP} results & & &  &   &   &    &      &  &    &     &  &    &  & &\\
				\noalign{\smallskip}
			\hline
				\noalign{\smallskip}
			2M17311037$-$2957146 &  4527   &  1.55   &   1.91   &   $-$0.31  &    $+$0.91  &  $+$0.27  &   $+$0.26  &    $+$0.40  &  $+$0.26  &   $+$0.49 &  $+$0.27  &   $+$0.15  &   $-$1.10  &   $+$0.02  &   $-$0.07  \\ 
			2M17305949$-$3000019 &  4445   &  1.07   &   2.51   &   $-$0.44  &    $+$0.87  &  $+$0.28  &   $+$0.19  &    $+$0.09  &  $+$0.26  &   $+$0.71 &  $+$0.20  &   $-$0.04  &   $-$1.26  &   $+$0.01  &   $-$0.31  \\ 
			2M17305774$-$2957404 &  4492   &  1.32   &   2.34   &   $-$0.30  &    $+$1.03  &  $+$0.20  &   $+$0.23  &    $+$0.47  &  $+$0.23  &   $+$0.41 &  $+$0.29  &   $+$0.13  &   $-$1.13  &   $+$0.03  &   $+$0.07  \\  
			2M17310585$-$2958354 &  4728   &  1.27   &   2.48   &   $-$0.55  &    $+$1.02  &  $+$0.25  &   $+$0.16  &    $+$0.13  &  $+$0.22  &   $+$0.56 &  $+$0.08  &   $+$0.03  &   $-$1.22  &   $-$0.01  &   $-$0.15  \\ 
			2M17310541$-$2958416 &  4200   &  1.10   &   1.88   &   $-$0.25  &    $+$0.24  &  $+$0.30  &   $+$0.31  &    $-$0.05  &  $+$0.26  &   $+$0.29 &  $+$0.28  &   $-$0.01  &   $-$1.14  &   $+$0.01  &   $-$0.08  \\ 
			2M17311196$-$2958587 &  4476   &  1.51   &   2.29   &   $-$0.34  &    $+$1.05  &  $+$0.20  &   $+$0.21  &    $+$0.67  &  $+$0.22  &   $+$0.28 &  $+$0.31  &   $+$0.20  &   $-$1.10  &   $+$0.06  &   $+$0.18  \\  
			2M17310703$-$2959376 &  4498   &  1.31   &   2.46   &   $-$0.46  &    $+$1.12  &  $+$0.13  &   $+$0.13  &    $+$0.80  &  $+$0.16  &   $+$0.38 &  $+$0.28  &   $+$0.23  &   $-$1.08  &   $+$0.01  &   $+$0.22  \\  
			2M17310020$-$2959415 &  4576   &  1.43   &   2.36   &   $-$0.44  &    $+$1.09  &  $+$0.17  &   $+$0.16  &    $+$0.68  &  $+$0.19  &   $+$0.36 &  $+$0.29  &   $+$0.17  &   $-$1.06  &   $+$0.01  &   $+$0.09  \\  
			2M17310045$-$2959502 &  4357   &  1.06   &   2.36   &   $-$0.37  &    $+$0.34  &  $+$0.30  &   $+$0.24  &    $-$0.14  &  $+$0.29  &   $+$0.55 &  $+$0.17  &   $-$0.14  &   $-$1.30  &   $-$0.12  &   $-$0.27  \\ 
			2M17304363$-$2954441 &  4817   &  1.53   &   2.39   &   $-$0.18  &    $+$0.80  &  $+$0.40  &   $+$0.20  &    $+$0.24  &  $+$0.26  &   $+$0.26 &  $+$0.14  &   $+$0.06  &   $-$1.16  &   $-$0.09  &   $+$0.06  \\   
				\noalign{\smallskip}
			\hline
				\noalign{\smallskip}
			Median & ... & ... & ... & $-$0.35 & $+$0.97 &   $+$0.26  &    $+$0.21    & $+$0.33  &  $+$0.24 &   $+$0.40   &   $+$0.28  &   $+$0.10  &   $-$1.13   &   $-$0.01    &  $+$0.01  \\
			\bf{Mean} & ... & ... & ...    & $-$0.36 & $+$0.85 &   $+$0.25  &    $+$0.21    & $+$0.33  &  $+$0.23 &   $+$0.43   &   $+$0.23  &   $+$0.08  &   $-$1.16   &   $-$0.01    &  $-$0.03  \\
			1$\sigma$ & ... & ... & ...   & 0.09 & 0.26 &   0.06  &    0.05    & 0.33  &  0.03 &   0.14   &   0.07  &   0.11  &   0.08   &   0.04    &  0.18 \\ 
				\noalign{\smallskip}
			\hline
		\end{tabular}\label{table3}
		\end{center}
	\raggedright{{\bf Note:} The asterisk (\textbf{$^*$}) mark the chemical species whose line profiles are heavily blended by other features, and for the purpose of this paper we warn the reader to take it as upper limits. Reference solar chemical abundances are those from \citet{Asplund2005}, except for Ce II for which we have adopted the Solar abundances from \citet{Grevesse2015}. The mean elemental abundances, and 1$\sigma$ error, defined as (84$^{\rm th}$ percentile $-$ 16$^{\rm th}$ percentile)/2, the standard deviation (std), and star-to-star spread are shown for the full sample.}
\end{small}
\end{table*}   
			
{ The CAPOS spectra provide access to 26 chemical species. However, most of the atomic and molecular lines are very weak and/or heavily blended, in some cases too much so to produce reliable abundances in  the typical T$_{\rm eff}$, $\log$(\textit{g}) and metallicity ranges of our sample. For this reason, and after a careful visual inspection of all our spectra, we provide reliable abundance determinations for twelve selected chemical species, belonging to the iron-peak (Fe, Ni), odd-Z (Al), light- (C, N), $\alpha$- (O, Mg, Si, S, Ca and Ti), and \textit{s}-process (Ce) elements, which are listed in Table \ref{table3}.}

{  We did not include Na in our analysis, which is a classical species to characterize MPs in GCs, as this relies on two atomic lines in the \textit{H}-band (Na I: 1.6363 $\mu$m and 1.6388 $\mu$m), which are unfortunately too weak and strongly blended by telluric features, and thus not able to produce reliable [Na/Fe] abundance ratios in HP~1. For this reason, in investigating MPs in HP 1, we emphasize C, N, O, Mg and Al, which are also typical chemical tracers of MPs in GCs \citep[see e.g.,][]{Meszaros2015, Ventura2016,  Meszaros_2020, Masseron2016, Pancino2017, Schiavon2017, Fernandez-Trincado2021a, Fernandez-Trincado2022b, Schiavon2024}.  }

{ It is important to note that two stars in our  sample (2M17304363$-$2954441 and 2M17310585$-$2958354) show no reliable C$^{12}$N$^{14}$ lines, therefore, it was not possible to determine their [N/Fe] abundance ratios. For these particular cases, we adopted the same prescription as described in \cite{Simpson_2019MNRAS.490..741S}, and evaluated the impact of the unknown [N/Fe] abundance ratios on [Fe/H] and [C,O/Fe], considering that carbon and oxygen abundance ratios are derived from the molecular equilibria that exist in the stellar atmosphere between C$^{12}$N$^{14}$, C$^{12}$O$^{16}$and O$^{16}$H. Thus, we assumed values of  $[{\rm N/Fe}]\in\{-1,-0.5,0.0,+0.5,+1.0\}$ and determined for each combination the corresponding [Fe/H] and [C,O/Fe] abundance ratios. The overall effect of the unknown [N/Fe] for these stars is relatively small: varying N through $-1.0 <$ [N/Fe] $<+1.0$ registers variations in $\Delta$[C/Fe]$<0.2$ dex, $\Delta$[O/Fe]$<0.2$ dex, and $\Delta$[Fe/H]$<0.1$ dex.  
}

{ For [S/Fe] abundance ratios, we find that  most of the S I lines are very weak and heavily blended by other features, in the typical T$_{\rm eff}$, $\log$(\textit{g}) and metallicity range of our sample, leading to unreliable [S/Fe] abundances. Therefere, the only listed [S/Fe] abundance ratios in Table \ref{table3} should be taken as upper limits. }

{ As the abundance values are sensitive to all of the atmospheric parameters, depending on the  chemical species, to estimate their uncertainties, we have varied the main atmospheric parameters one at a time and  computed the abundances for all species for each of these possibilities for one star, 2M17305949$-$3000019, and calculated the typical uncertainty as $\sigma_{\rm total}$, following the same methodology outlined in \cite{Fernandez-Trincado2020a}, and repeated in Eq. \ref{eq2} for guidance. The total uncertainties, $\sigma_{\rm total}$, is defined as: }
		
\begin{equation}\label{eq2}
\sigma^{2}_{\rm total}  = \sigma^2_{\rm [X/H], T_{\rm eff}}    + \sigma^2_{\rm [X/H],{\rm log} (\textit{g})} + \sigma^2_{\rm [X/H],\xi_t}  + \sigma^2_{\rm mean}  
\end{equation}

\noindent
{ where $\sigma^2_{mean}$ is calculated using the standard deviation derived from the different \texttt{BACCHUS} abundances of the different lines (line-by-line variation) for each element. The values of $\sigma^2_{[X/H], T_{\rm eff}}$, $\sigma^2_{\rm [X/H],{\rm log} (\textit{g})}$, and $\sigma^2_{\rm [X/H],\xi_t} $ are derived for the elements in 2M17305949$-$3000019 using the sensitivity values of $\pm100$ K for the temperature, $\pm0.3$ dex for log \textit{g}, and 0.05 km s$^{-1}$ for the microturbulent velocity ($\xi_{t}$). These values were chosen as they represent the typical conservative uncertainties in the atmospheric parameters for our sample. It is important to notice that we have have found that lower-(S/N) and/or higher-(S/N) produce similar results in their uncertainties, with negligible variations in the final $\sigma_{\rm total}$. Thus, the star 2M17305949$-$3000019 is a good representation of the HP~1 sample and it is reasonable to assume that the errors we find  reflect the typical cluster star analyzed in this study. Therefore, the uncertainties listed in Table \ref{table3} for this star are considered as representative of our HP~1 sample. 
}

\begin{figure*}
	\centering
	\includegraphics[scale=0.19]{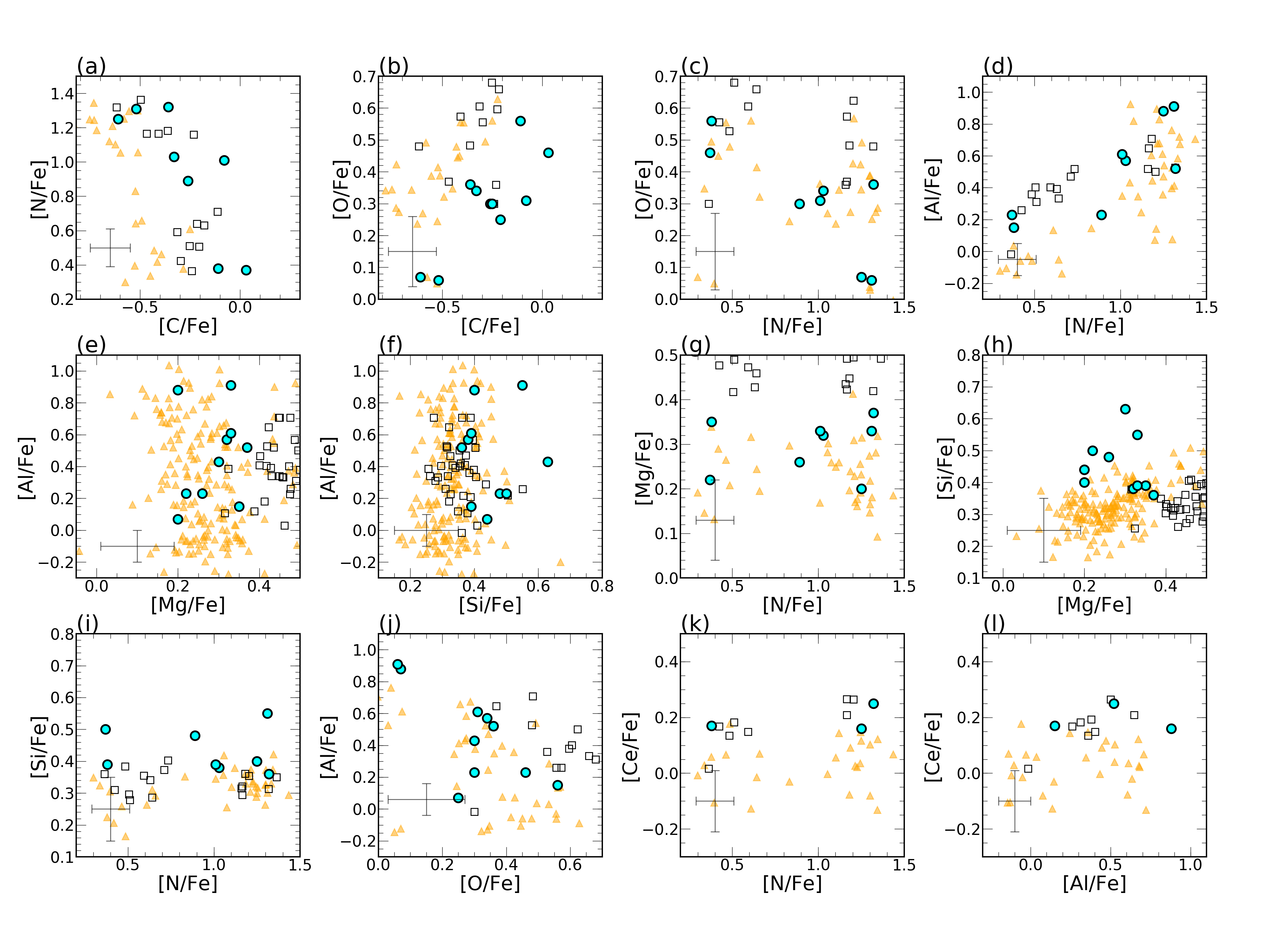}
	\caption{ Combined light-, odd-Z, $\alpha$- and \textit{s}-process elements from our \texttt{BACCHUS} results listed in Table \ref{table3}. \textit{Panels} (a)--(l): [C/Fe]--[N/Fe], [C/Fe]--[O/Fe], [N/Fe]--[O/Fe], [N/Fe]--[Al/Fe], [Mg/Fe]--[Al/Fe], [Si/Fe]--[Al/Fe], [N/Fe]--[Mg/Fe], [Mg/Fe]--[Si/Fe], [N/Fe]--[Si/Fe], [O/Fe]--[Al/Fe], [N/Fe]--[Ce/Fe], and [Al/Fe]--[Ce/Fe] distributions for HP~1 (cyan circles), in comparison with the others APOGEE-2 GCs examined by \citet{Meszaros_2020}: NGC~288 (black empty squares) and NGC~5904  (orange triangles). The error bars mark the typical uncertainty ($\sigma_{\rm total}$) listed in Table \ref{table3}. }
	\label{figure5}
\end{figure*}

\section{Results and discussion}\label{Results}

{ The present study adds a substantial contribution to the chemical characterization of HP~1, including detailed abundances for 12 chemical species derived from high S/N, high resolution near-IR spectra of 10 member giants. In the only other high resolution spectroscopic study, Barbuy (2006,2016) analyzed a total of eight red giants in this cluster, but using optical spectra. Our sample of likely cluster members increases this number, and is thus the largest sample yet analyzed spectroscopically,  with the added benefit of observing in the near-IR to help mitigate this cluster's high reddening, allowing us to examine abundance variations  within the cluster. 

Figure \ref{figure5} grapically illustrates the chemical behaviour of HP~1 stars as well as comparing them to two other GCs at similar metallicity (NGC~5904 and NGC~288) taken from \citep{Meszaros_2020}, whose data set has been carefully examined with the same code and similar methodology as adopted in this work. We avoid any comparison with GCs only based on the \texttt{ASPCAP} APOGEE-2 pipeline as they may exhibit larger systematic offsets \citep[see e.g.,][]{Jonsson_2018, Holtzman2018, Nataf2019}. Thus, the listed \texttt{ASPCAP} values in Table \ref{table3} are only for reference, and are not considered in our analysis, as our BACCHUS results are likely more precise that  \texttt{ASPCAP}. 
} 

\subsection{The Fe-peak elements: Fe and Ni}

{ We measure a mean metallicity of $\langle$[Fe/H]$\rangle = -1.15\pm 0.03$ (standard error of the mean), with a dispersion of  0.08 $\pm$ 0.02 dex. This value is $< 0.1 $ dex more metal poor than the mean  [Fe/H]$= -1.06$ reported by \citet{Barbuy2016A&A...591A..53B}, used as our initial metallicity estimate. This discrepancy is within the relative errors. Thus, we conclude that our mean [Fe/H] is in good agreement with the value reported in \citet{Barbuy2016A&A...591A..53B}, and HP~1 is a cluster with an intermediate metallicity, [Fe/H]$ \sim -1.15$.
Note that the mean ASPCAP value is very similar, -1.16, and also has the same dispersion. Geisler et al. (2021) derived a mean of $-1.20\pm 0.10$ from the ASPCAP DR16 Fe abundances for the same sample, while Geisler et al. (in prep.) find 
$-1.23\pm 0.07$ from DR 17.
However, both of these values employed a small negative correction to the ASPCAP metallicities of stars with high [N/Fe], which is indeed the case for most of our sample, to correct for ASPCAP issues for such 
second generation stars. 
Schiavon et al. (2023) finds -1.21 for a somewhat different DR17 sample, and does not include a metallicity correction.
We finally note that the Harris (2010) catalog value for this cluster is -1, based on lower quality data.
}

{ Our observed star-to-star [Fe/H] spread of  0.25 dex is quite similar to our measurement uncertainty (see Table \ref{table3}), indicating that no significant metallicity variation is detected. }

{ Regarding the other iron-peak element we examined, nickel (Ni) is on average slightly super-solar ($\langle$[Ni/Fe]$\rangle =+0.03\pm 0.03$) with a very small dispersion, $<0.06$ dex, and a very small star-to-star spread ($\sim$0.11 dex),  within our typical uncertainties. Therefore, we also do not detect a significant spread in this element. 
}

\subsection{The odd-Z element: Al}

{ We find that HP~1 exhibits a mean aluminium enrichment of $\langle$[Al/Fe]$\rangle =+0.46\pm 0.29$, with a large variation in [Al/Fe] which is far beyond the typical errors. Thus, we report a strong star-to-star [Al/Fe]  spread of $\sim0.84$ dex in HP~1, which is expected for GCs at similar metallicity due to MPs \citep[see e.g.,][]{Pancino2017, Masseron2019, Meszaros_2020}. }

{ Figures \ref{figure5}(\textit{e}--\textit{f}) reveals the strong star-to-star [Al/Fe] spread observed in HP~1, which is very similar to NGC~5904 (M~5) and NGC~288. It is clear that the extended distribution of Al is clearly larger that the estimated errors of [Al/Fe] in HP~1, and suggests an astrophysical origin, which is likely the result of the past activation of the Mg-Al cycle nuclear fusion process at, possibly, the early stages of  asymptotic giant branch star (AGB) evolution \citep{Meszaros2015, Ventura2016}. However, figures \ref{figure5}(\textit{e}--\textit{f}; \textit{h}), do not reveal any significant/clear Mg--Al/Si--Al (anti)correlations in HP~1.
Similar behavior is also seen in the comparison GCs.

If the Mg--Al fusion cycle is involved, which converts Mg into Al, it apparently does not completely process the material through the entire Mg--Al cycle, suggesting that the observed large star-to-star [Al/Fe] spread  may also be due in part to a diversity in the process of stellar chemical feedback and star formation in these clusters. Note that we have not been able to find any dependence of the Al abundance on effective temperature or evolutionary status (RGB or AGB), ruling out the presence of any possible analysis bias.}

\begin{figure}[h]
	\centering
	\includegraphics[scale=0.21]{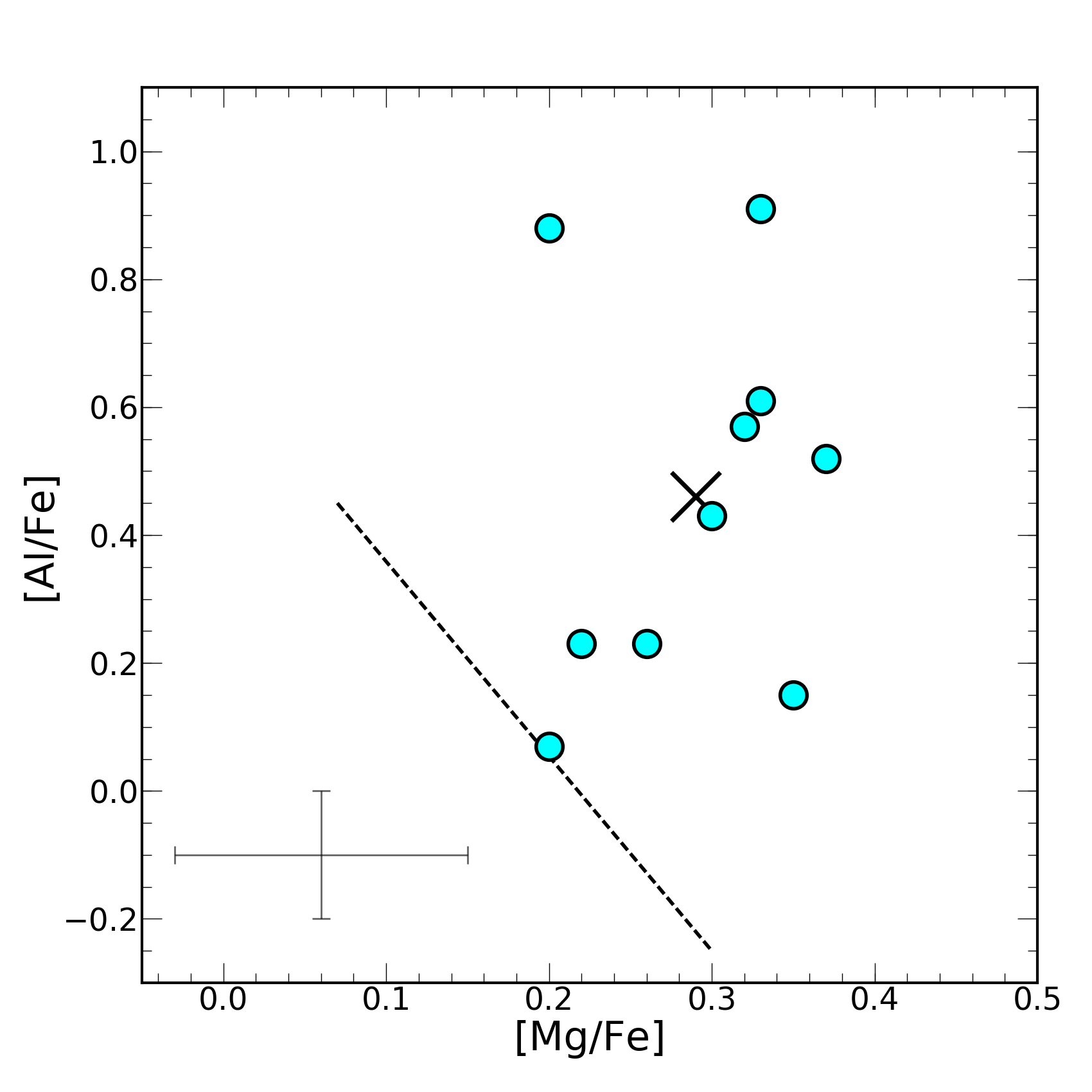}
	\caption{ [Mg/Fe] -- [Al/Fe] distribution for HP~1 stars (cyan symbols). The black ``X'' symbols show the average [Mg/Fe] and [Al/Fe] abundance ratios of HP~1, from our \texttt{BACCHUS} values listed in Table \ref{table3}. The dashed black line taken from \citet{belokurov2024situ}, separate the  region dominated by \textit{in-situ} GCs (above the line) from accreted GC (below the line).}
\label{belokurov}
\end{figure}

{ Figure \ref{belokurov} reveals that the [Al/Fe] vs. [Mg/Fe] abundance ratios of HP~1 stars fall clearly in the region dominated by \textit{in-situ} GCs as defined by \citet{belokurov2024situ}, supporting our claim that HP~1  likely formed \textit{in-situ}, as expected for a BGC. }

\subsection{The $\alpha$-elements: O, Mg, Si, Ca and Ti}

{ Table \ref{table3} reveals that HP~1 exhibits a considerable $\alpha$-element enhancement, with mean values ranging from $+$0.20 to $+$0.45 ([O/Fe], [Mg/Fe], [Si/Fe], [Ca/Fe] and [Ti/Fe]), and in reasonable agreement with GCs of similar metallicity such as NGC~5904 and NGC~288, with small star-to-star spread within our typical uncertainties, except for [O/Fe], [Si/Fe] and [Ca/Fe], which show a star-to-star spread that exceeds the observational uncertainties. [O/Fe] is the only $\alpha$-element with an very extended and significant spread, $>0.5$ dex, which is clearly (anti)correlated with light-elements (C and N) as shown in Figures \ref{figure5}(\textit{b}--\textit{c}).} 
	
{ It is also expected that Al is correlated with elements enhanced by proton-capture reactions (nitrogen) as demonstrate in Figure \ref{figure5}(\textit{d}), and anticorrelated with those depleted in H-burning at high temperature (oxygen) as revealed in Figure \ref{figure5}(\textit{j}). Therefore, Al and O represent  more robust  indicators for the prevalence of the MPs phenomenon in HP~1. 
} 

\begin{figure}[h]
	\centering
	\includegraphics[scale=0.15]{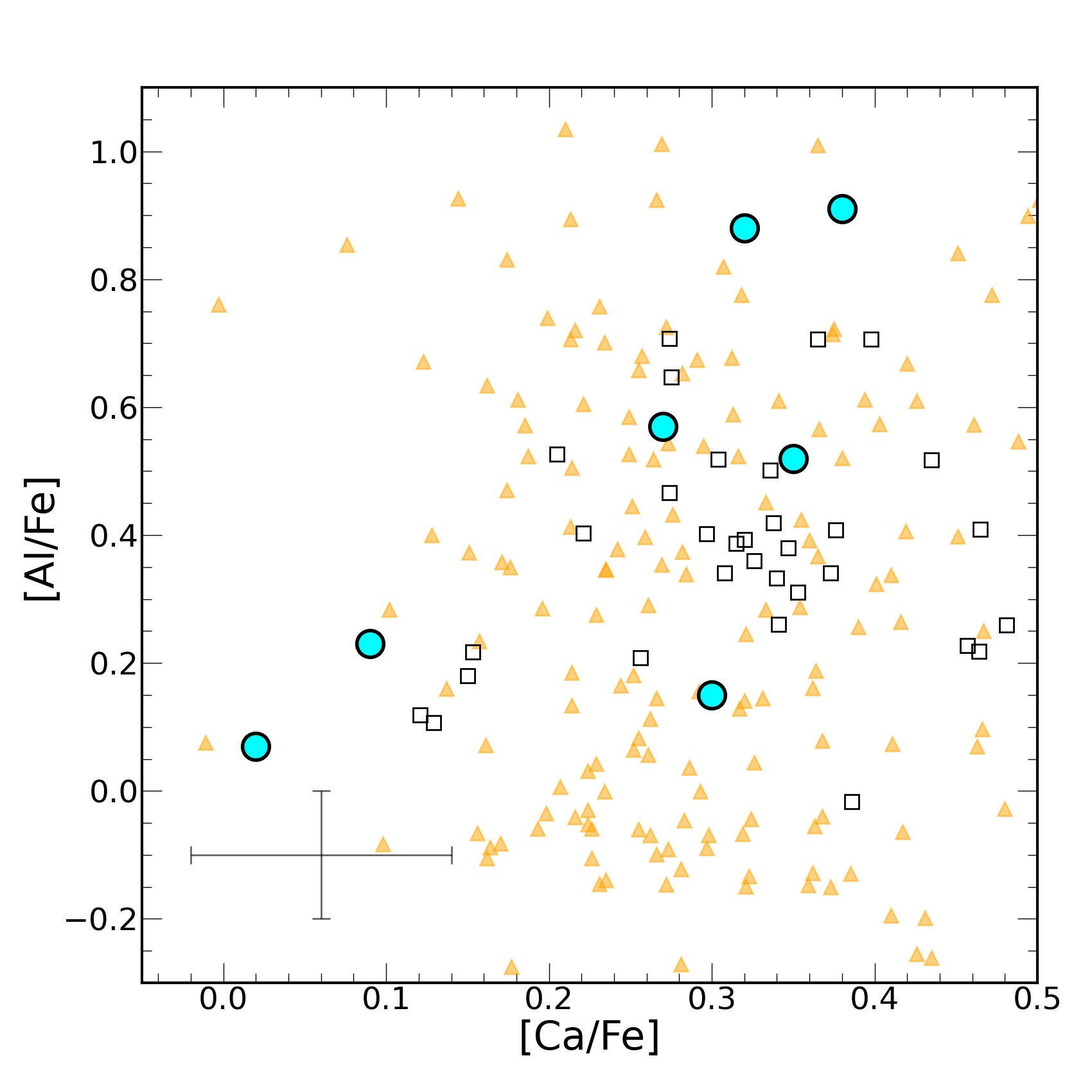}
	\caption{ [Ca/Fe] -- [Al/Fe] distribution for HP~1 stars (cyan symbols) and our two comparison GCs.}
\label{figure7}
\end{figure}

{ 
We also find a [Ca/Fe] abundance ratio scatter in HP~1, 0.36 dex, which is larger that the typical internal uncertainties. 
Figure \ref{figure7}
reveals an intriguing 
correlation between [Ca/Fe] and [Al/Fe], which has not been previously seen in any bulge GC to our knowledge. 
The significance of this correlation is 0.74 (Pearson's test) with a \textit{p}-value of 0.05, and 0.75 (Spearman's test) with a \textit{p}-value of 0.05, 
suggesting the correlation is indeed significant.
However, we note that the sample is small (only 7 stars with measurable [Ca/Fe] abundances ratios). 
Moreover, 2 of the stars are noted as having blending issues in Al and 3 in Ca. Thus, more observations will be needed to substantiate the reality of such a correlation. In addition, neither of our comparison GCs displays any such correlation, making our result more suspicious.
Astrophyically, 
the Al spread of course is due to  the outcome of proton-capture reactions at very high temperatures \citep{Carretta2021}. However, while a single core-collapse super-nova might be enough to produce the claimed star-to-star spread in [Ca/Fe], fast rotating massive stars \citep[FRMS;][]{Decressin2007} or massive AGB stars have been proposed to produce the observed pattern in Al \citep{Carretta2010}, thus requiring multiple astrophyical processes to account for any correlation.  Further research is clearly required to corroborate this intriguing possibility.  
}

{ In general, we find that HP~1 exhibits average [Mg, O/Fe] abundance ratios which are in good agreement with those obtained by \cite{Barbuy2016A&A...591A..53B}, which range from $+$0.3 $\lesssim$ [O,Mg/Fe] $\lesssim$ $+$0.5. 
Of course, the mean will depend on the MP nature of each sample, given that both of these elements are affected by MPs. 

Our data yield a mean [Si/Fe] = +0.45, with
some
stars having enrichments as large as $+0.63$.
The similar-sized sample analyzed by \cite{Barbuy2016A&A...591A..53B} is limited to $+0.15$ $\lesssim$ [Si/Fe] $\lesssim +0.35$. Indeed, our [Si/Fe] abundances appear high compared to our comparison GCs of similar metallicity in Figure 5 as well.
We note our average [Ca/Fe] and [Ti/Fe] abundance ratios are moderately enhanced compared to that measured by \cite{Barbuy2016A&A...591A..53B}, which reports values in the range of $-$0.04 $\lesssim$ [Ca,Ti/Fe]$\lesssim$ $+0.28$.}

{ 
	In order to validate the reliability of the observed [O/Fe] trends in HP~1, and to examine any possible bias introduced by the stellar parameters in the interpretation and discussion of our results, we have applied the Pearson's and Spearman's correlation tests to the T$_{\rm eff}$--[O/Fe] and [N/Fe]--[O/Fe] planes, and validate our results by applying a bootstrap resampling with 100,000 realizations. The Pearson's and Spearman's correlation coefficients range from $-0.22 -- -0.30$ with \textit{p}-value $>0.3$ indicating that there is a null trend between [O/Fe] and T$_{\rm eff}$; whilst the Pearson's and Spearman's correlation coefficients range from $-0.52$ to $-0.93$ with a low \textit{p}$-$value ($<$0.02) indicating that the observed [N/Fe]--[O/Fe] anticorrelation is unlikely due to random chance. 
}

\subsection{The light-elements: C and N}

{ HP~1 exhibits a high enrichment in nitrogen, with a mean $\langle$[N/Fe]$\rangle = +0.94\pm0.43$, and a large star-to-star spread of $+0.95$ dex, typical of MPs. HP~1 reveals a statistically significant correlation in C--O, as shown in Figure \ref{figure5}(\textit{b}), and clear anti-correlations in C--N (Figure \ref{figure5}--\textit{a}) and N--O (Figure \ref{figure5}--\textit{c}). These patterns are typical of GCs and are  attributed to the prevalence of the MPs phenomenon \citep[see e.g.,][and references therein]{Schiavon2017, Schiavon2024}. Almost all the stars examined in HP~1 are enriched in nitrogen (except two stars with [N/Fe]$<+0.4$). The chemical trends of [N/Fe] is very similar to those observed in NGC~5904 and NGC~288 (see Figure \ref{figure5}), but with (at least) two groups of stars, likely compatible with a first generation ([N/Fe]$\lesssim +0.7$, and a second generation ([N/Fe]$\gtrsim+0.7$) (Geisler et al. 2021). This study reveals that a significant fraction of the stars with enhanced [N/Fe] abundances well above Galactic levels ([N/Fe]$\gtrsim+0.7$) populate HP~1, a feature that is typical of stars in BGCs such as NGC~6380 \citep{Fernandez-Trincado2021d}, and a clear indication of MPs \citep[see, e.g.,][]{Schiavon2017, Geisler_2021_2021A&A...652A.157G, Schiavon2024}. }

{ Here we also report for the first time [C/Fe] abundance ratios for this cluster, which are in the range $-$0.61 $<$ [C/Fe] $< +$0.03, while \cite{Barbuy2016A&A...591A..53B} reported only an upper limit  for [C/Fe] $<$ 0.}

{ 
The CNO trends we find are in agreement with the output of the hot CNO cycle \citep{Wiescher2010} that predicts lower C and O and higher N abundances for the material that goes through it. However, the CNO trends we find cannot discriminate between the different scenarios that have been proposed to explain the MPs phenomenon, and in particular they cannot discriminate which polluter is responsible.
}

\subsection{The \textit{s}-process element: Ce}

{ We find a mean $\langle$[Ce/Fe]$\rangle =+0.19\pm 0.03$ from three stars in HP~1, which is moderately overabundant compared to the Sun, and similar to the Ce levels observed in other GCs such as NGC~288 (see Figure \ref{figure5}(\textit{l})). Thus, we believe that this  moderate enrichment in [Ce/Fe] is likely produced by different progenitors, possibly by pollution of this cluster by low mass AGB stars \citep{Fernandez-Trincado2021a, Fernandez-Trincado2022b} after the cluster formed. Unfortunately, with only one neutron-capture element measured in HP~1 for only three stars with reliable Ce II line profiles, it is not possible to firmly assign the nucleosynthetic origins of these stars.
}

\section{Concluding remarks}\label{summary}

{  We present a detailed elemental-abundance analysis using the BACCHUS package for ten stars belonging to the BGC HP~1, obtained from high S/N, high resolution near-IR spectra with APOGEE as part of the CAPOS survey. We examined twelve chemical species belonging to the  light- (C and N), $\alpha$- (O, Mg, Si, S, Ca, and Ti), iron-peak (Fe and Ni), odd-Z (Al), and \textit{s}-process (Ce) elements. Overall, the chemical species examined so far in HP~1 are in agreement with other Galactic GCs at similar metallicity \citep[e.g.,][]{Masseron2019, Meszaros2015, Schiavon2017, Meszaros_2020, Geisler_2021_2021A&A...652A.157G, Schiavon2024}. The main conclusions of this paper are the following:}

\begin{itemize}
	
	\item { HP~1 exhibits a mean $\langle$[Fe/H]$\rangle = -1.15\pm0.03$, with a star-to-star [Fe/H] spread, $\sim$0.25 dex, which  is quite similar to our measurement uncertainty. Thus, no significant metallicity variation was detected. Our reported average [Fe/H] is in good agreement with the metallicity estimated by \citet{Barbuy2016A&A...591A..53B}.}

    \item { 
    $[\alpha /Fe]$ ratios are enhanced, as expected, with [Si/Fe] being especially high, substantially higher than found by \citet{Barbuy2016A&A...591A..53B}.
    }
	
	\item {  The typical Mg-Al anti-correlation is not observed, which is in line with other GCs at similar metallicity \citep[see, e.g.,][]{Meszaros_2020}. However, a significant star-to-star [Al/Fe] spread, $0.84$ dex, was identified, which (anti)correlates with [O/Fe] and [N/Fe]. We conclude that the O--Al anticorrelation and  N-Al correlations represent robust indicators for the prevalence of the MPs phenomenon in HP~1.} 
	
	\item { We find for the first time a significant variation in C, N, and O, with a clear [C/Fe]--[N/Fe] and [N/Fe]--[O/Fe] anticorrelation, and [C/Fe]--[O/Fe] correlation, and a significant spread in nitrogen ($>0.95$ dex), carbon ($>0.64$ dex), and oxygen ($>0.5$), as expected from MPs. }
	
	\item {  HP~1 also exhibits a modest star-to-star [Ca/Fe] spread, $+0.36$,  that exceeds the observational uncertainties. Furthermore, [Ca/Fe] correlates with [Al/Fe]. However, we note that the sample size is quite small and observationally limited. This finding, observed for the first time in a bulge GC, if confirmed, makes HP~1  an especially interesting representative.
	}
	
	\item {  Unfortunately, there are too few stars with reliable measurements of [Ce/Fe] and [S/Fe] abundance ratios to provide reliable conclusions regarding these chemical species. 
	}
	
	\item { HP~1 hosts a significant population of nitrogen-enriched stars indicating the prevalence of the MPs phenomenon in this cluster. Furthermore, the high nitrogen enrichment of HP~1 makes this cluster a potential progenitor of the unusual nitrogen-enhanced bulge field stars identified in the inner Galaxy, peaking at [Fe/H]$\sim-1.0$ \citep[e.g.,][]{Schiavon2017Nitrogen, Fernandez-Trincado2022Nitrogen}.
	}

	\item { The mean $\langle$[Mg/Fe]$\rangle = +0.29$ and $\langle$[Al/Fe]$\rangle = +0.46$ measured in this work place HP~1 into the region dominated by \textit{in-situ} GCs \citep[see e.g.,][]{belokurov2024situ}, supporting the \textit{in-situ} nature of HP~1. Despite the observed very large scatter in [Al/Fe], all  stars fall in the \textit{in-situ} domain. 
}

	
\end{itemize}

{ In this work, we obtained one of the first detailed chemical analyses of one of the oldest \textit{in-situ} GCs. 
Further investigation, including detailed research on elements with distinct nucleosynthesis processes and comparison with a larger sample of HP 1 stars as well as similar clusters, should provide valuable insights into the unique evolutionary path of this GC  in the bulge, and contribute to our broader understanding of GCs formation and evolution in the MW in general, which is the goal of the CAPOS survey.
Finally, these results underscore the complex nature of chemical abundance patterns in GCs and highlight the importance of comprehensive NIR spectroscopic studies in unraveling the chemical signature for the prevalence of the MPs phenomenon across the entire volume of parameter space covered by CAPOS.}

\begin{acknowledgements}
The first author gratefully acknowledges the support provided by the National Agency for Research and Development (ANID)/CONICYT-PFCHA/DOCTORADO NACIONAL/2017-21171231.
S.V. gratefully acknowledges the support provided by Fondecyt Regular n. 1220264, and by the ANID BASAL project ACE210002.
{ D.G. gratefully acknowledges the support provided by Fondecyt regular n. 1220264.
D.G. also acknowledges financial support from the Direcci\'on de Investigaci\'on y Desarrollo de
la Universidad de La Serena through the Programa de Incentivo a la Investigaci\'on de
Acad\'emicos (PIA-DIDULS).
J.G.F-T gratefully acknowledges the grants support provided by ANID Fondecyt Iniciaci\'on No. 11220340, ANID Fondecyt Postdoc No. 3230001, and from the Joint Committee ESO-Government of Chile under the agreement 2023 ORP 062/2023.}
\end{acknowledgements}


\begin{thebibliography}{72}
	\expandafter\ifx\csname natexlab\endcsname\relax\def\natexlab#1{#1}\fi
	
	\bibitem[{{Abdurro'uf} {et~al.}(2022){Abdurro'uf}, {Accetta}, {Aerts}, {Silva
			Aguirre}, {Ahumada}, {Ajgaonkar}, {Filiz Ak}, {Alam}, {Allende Prieto},
		{Almeida}, {Anders}, {Anderson}, {Andrews}, {Anguiano}, {Aquino-Ort{\'\i}z},
		{Arag{\'o}n-Salamanca}, {Argudo-Fern{\'a}ndez}, {Ata}, {Aubert},
		{Avila-Reese}, {Badenes}, {Barb{\'a}}, {Barger}, {Barrera-Ballesteros},
		{Beaton}, {Beers}, {Belfiore}, {Bender}, {Bernardi}, {Bershady}, {Beutler},
		{Bidin}, {Bird}, {Bizyaev}, {Blanc}, {Blanton}, {Boardman}, {Bolton},
		{Boquien}, {Borissova}, {Bovy}, {Brandt}, {Brown}, {Brownstein}, {Brusa},
		{Buchner}, {Bundy}, {Burchett}, {Bureau}, {Burgasser}, {Cabang}, {Campbell},
		{Cappellari}, {Carlberg}, {Wanderley}, {Carrera}, {Cash}, {Chen}, {Chen},
		{Cherinka}, {Chiappini}, {Choi}, {Chojnowski}, {Chung}, {Clerc}, {Cohen},
		{Comerford}, {Comparat}, {da Costa}, {Covey}, {Crane}, {Cruz-Gonzalez},
		{Culhane}, {Cunha}, {Dai}, {Damke}, {Darling}, {Davidson}, {Davies},
		{Dawson}, {De Lee}, {Diamond-Stanic}, {Cano-D{\'\i}az}, {S{\'a}nchez},
		{Donor}, {Duckworth}, {Dwelly}, {Eisenstein}, {Elsworth}, {Emsellem},
		{Eracleous}, {Escoffier}, {Fan}, {Farr}, {Feng}, {Fern{\'a}ndez-Trincado},
		{Feuillet}, {Filipp}, {Fillingham}, {Frinchaboy}, {Fromenteau}, {Galbany},
		{Garc{\'\i}a}, {Garc{\'\i}a-Hern{\'a}ndez}, {Ge}, {Geisler}, {Gelfand},
		{G{\'e}ron}, {Gibson}, {Goddy}, {Godoy-Rivera}, {Grabowski}, {Green},
		{Greener}, {Grier}, {Griffith}, {Guo}, {Guy}, {Hadjara}, {Harding},
		{Hasselquist}, {Hayes}, {Hearty}, {Hern{\'a}ndez}, {Hill}, {Hogg},
		{Holtzman}, {Horta}, {Hsieh}, {Hsu}, {Hsu}, {Huber}, {Huertas-Company},
		{Hutchinson}, {Hwang}, {Ibarra-Medel}, {Chitham}, {Ilha}, {Imig}, {Jaekle},
		{Jayasinghe}, {Ji}, {Johnson}, {Jones}, {J{\"o}nsson}, {Katkov}, {Khalatyan},
		{Kinemuchi}, {Kisku}, {Knapen}, {Kneib}, {Kollmeier}, {Kong}, {Kounkel},
		{Kreckel}, {Krishnarao}, {Lacerna}, {Lane}, {Langgin}, {Lavender}, {Law},
		{Lazarz}, {Leung}, {Leung}, {Lewis}, {Li}, {Li}, {Lian}, {Liang}, {Lin},
		{Lin}, {Lin}, {Lintott}, {Long}, {Longa-Pe{\~n}a}, {L{\'o}pez-Cob{\'a}},
		{Lu}, {Lundgren}, {Luo}, {Mackereth}, {de la Macorra}, {Mahadevan},
		{Majewski}, {Manchado}, {Mandeville}, {Maraston}, {Margalef-Bentabol},
		{Masseron}, {Masters}, {Mathur}, {McDermid}, {Mckay}, {Merloni},
		{Merrifield}, {Meszaros}, {Miglio}, {Di Mille}, {Minniti}, {Minsley}, \&
		{Monachesi}}]{Abdurro2022}
	{Abdurro'uf}, {Accetta}, K., {Aerts}, C., {et~al.} 2022, \apjs, 259, 35
	
	\bibitem[{{Asplund} {et~al.}(2005){Asplund}, {Grevesse}, \&
		{Sauval}}]{Asplund2005}
	{Asplund}, M., {Grevesse}, N., \& {Sauval}, A.~J. 2005, in Astronomical Society
	of the Pacific Conference Series, Vol. 336, Cosmic Abundances as Records of
	Stellar Evolution and Nucleosynthesis, ed. I.~{Barnes}, Thomas~G. \& F.~N.
	{Bash}, 25
	
	\bibitem[{{Barb{\'a}} {et~al.}(2019){Barb{\'a}}, {Minniti}, {Geisler},
		{Alonso-Garc{\'\i}a}, {Hempel}, {Monachesi}, {Arias}, \&
		{G{\'o}mez}}]{Barba_2019ApJ...870L..24B}
	{Barb{\'a}}, R.~H., {Minniti}, D., {Geisler}, D., {et~al.} 2019, \apjl, 870,
	L24
	
	\bibitem[{{Barbuy} {et~al.}(2016){Barbuy}, {Cantelli}, {Vemado}, {Ernandes},
		{Ortolani}, {Saviane}, {Bica}, {Minniti}, {Dias}, {Momany}, {Hill},
		{Zoccali}, \& {Siqueira-Mello}}]{Barbuy2016A&A...591A..53B}
	{Barbuy}, B., {Cantelli}, E., {Vemado}, A., {et~al.} 2016, \aap, 591, A53
	
	\bibitem[{{Bastian} {et~al.}(2013){Bastian}, {Lamers}, {de Mink}, {Longmore},
		{Goodwin}, \& {Gieles}}]{BASTIAN_2013MNRAS.436.2398B}
	{Bastian}, N., {Lamers}, H.~J.~G.~L.~M., {de Mink}, S.~E., {et~al.} 2013,
	\mnras, 436, 2398
	
	\bibitem[{{Bastian} \& {Lardo}(2018)}]{Bastian2018}
	{Bastian}, N. \& {Lardo}, C. 2018, \araa, 56, 83
	
	\bibitem[{Belokurov \& Kravtsov(2024)}]{belokurov2024situ}
	Belokurov, V. \& Kravtsov, A. 2024, Monthly Notices of the Royal Astronomical
	Society, 528, 3198
	
	\bibitem[{Bica {et~al.}(2016)Bica, Ortolani, \&
		Barbuy}]{bica_ortolani_barbuy_2016}
	Bica, E., Ortolani, S., \& Barbuy, B. 2016, Publications of the Astronomical
	Society of Australia, 33, e028
	
	\bibitem[{{Bica} {et~al.}(2016){Bica}, {Ortolani}, \& {Barbuy}}]{bica2016}
	{Bica}, E., {Ortolani}, S., \& {Barbuy}, B. 2016, \pasa, 33, e028
	
	\bibitem[{{Blanton} {et~al.}(2017){Blanton}, {Bershady}, {Abolfathi},
		{Albareti}, {Allende Prieto}, {Almeida}, {Alonso-Garc{\'\i}a}, {Anders},
		{Anderson}, {Andrews}, {Aquino-Ort{\'\i}z}, {Arag{\'o}n-Salamanca},
		{Argudo-Fern{\'a}ndez}, {Armengaud}, {Aubourg}, {Avila-Reese}, {Badenes},
		{Bailey}, {Barger}, {Barrera-Ballesteros}, {Bartosz}, {Bates}, {Baumgarten},
		{Bautista}, {Beaton}, {Beers}, {Belfiore}, {Bender}, {Berlind}, {Bernardi},
		{Beutler}, {Bird}, {Bizyaev}, {Blanc}, {Blomqvist}, {Bolton}, {Boquien},
		{Borissova}, {van den Bosch}, {Bovy}, {Brandt}, {Brinkmann}, {Brownstein},
		{Bundy}, {Burgasser}, {Burtin}, {Busca}, {Cappellari}, {Delgado Carigi},
		{Carlberg}, {Carnero Rosell}, {Carrera}, {Chanover}, {Cherinka}, {Cheung},
		{G{\'o}mez Maqueo Chew}, {Chiappini}, {Choi}, {Chojnowski}, {Chuang},
		{Chung}, {Cirolini}, {Clerc}, {Cohen}, {Comparat}, {da Costa}, {Cousinou},
		{Covey}, {Crane}, {Croft}, {Cruz-Gonzalez}, {Garrido Cuadra}, {Cunha},
		{Damke}, {Darling}, {Davies}, {Dawson}, {de la Macorra}, {Dell'Agli}, {De
			Lee}, {Delubac}, {Di Mille}, {Diamond-Stanic}, {Cano-D{\'\i}az}, {Donor},
		{Downes}, {Drory}, {du Mas des Bourboux}, {Duckworth}, {Dwelly}, {Dyer},
		{Ebelke}, {Eigenbrot}, {Eisenstein}, {Emsellem}, {Eracleous}, {Escoffier},
		{Evans}, {Fan}, {Fern{\'a}ndez-Alvar}, {Fernandez-Trincado}, {Feuillet},
		{Finoguenov}, {Fleming}, {Font-Ribera}, {Fredrickson}, {Freischlad},
		{Frinchaboy}, {Fuentes}, {Galbany}, {Garcia-Dias},
		{Garc{\'\i}a-Hern{\'a}ndez}, {Gaulme}, {Geisler}, {Gelfand},
		{Gil-Mar{\'\i}n}, {Gillespie}, {Goddard}, {Gonzalez-Perez}, {Grabowski},
		{Green}, {Grier}, {Gunn}, {Guo}, {Guy}, {Hagen}, {Hahn}, {Hall}, {Harding},
		{Hasselquist}, {Hawley}, {Hearty}, {Gonzalez Hern{\'a}ndez}, {Ho}, {Hogg},
		{Holley-Bockelmann}, {Holtzman}, {Holzer}, {Huehnerhoff}, {Hutchinson},
		{Hwang}, {Ibarra-Medel}, {da Silva Ilha}, {Ivans}, {Ivory}, {Jackson},
		{Jensen}, {Johnson}, {Jones}, {J{\"o}nsson}, {Jullo}, {Kamble}, {Kinemuchi},
		{Kirkby}, {Kitaura}, {Klaene}, {Knapp}, {Kneib}, {Kollmeier}, {Lacerna},
		{Lane}, {Lang}, {Law}, {Lazarz}, {Lee}, {Le Goff}, {Liang}, {Li}, {Li},
		{Lian}, {Lima}, {Lin}, {Lin}, {Bertran de Lis}, {Liu}, {de Icaza Lizaola},
		{Long}, {Lucatello}, {Lundgren}, {MacDonald}, {Deconto Machado}, {MacLeod},
		{Mahadevan}, {Geimba Maia}, {Maiolino}, {Majewski}, {Malanushenko},
		{Malanushenko}, {Manchado}, {Mao}, {Maraston}, {Marques-Chaves}, {Masseron},
		{Masters}, {McBride}, {McDermid}, {McGrath}, {McGreer}, {Medina Pe{\~n}a},
		{Melendez}, {Merloni}, {Merrifield}, {Meszaros}, {Meza}, {Minchev},
		{Minniti}, {Miyaji}, {More}, {Mulchaey}, {M{\"u}ller-S{\'a}nchez}, {Muna},
		{Munoz}, {Myers}, {Nair}, {Nandra}, {Correa do Nascimento}, {Negrete},
		{Ness}, {Newman}, {Nichol}, {Nidever}, {Nitschelm}, {Ntelis}, {O'Connell},
		{Oelkers}, {Oravetz}, {Oravetz}, {Pace}, {Padilla}, {Palanque-Delabrouille},
		{Alonso Palicio}, {Pan}, {Parejko}, {Parikh}, {P{\^a}ris}, {Park}, {Patten},
		{Peirani}, {Pellejero-Ibanez}, {Penny}, {Percival}, {Perez-Fournon},
		{Petitjean}, {Pieri}, {Pinsonneault}, {Pisani}, {Poleski}, {Prada},
		{Prakash}, {Queiroz}, {Raddick}, {Raichoor}, {Barboza Rembold}, {Richstein},
		{Riffel}, {Riffel}, {Rix}, {Robin}, {Rockosi}, {Rodr{\'\i}guez-Torres},
		{Roman-Lopes}, {Rom{\'a}n-Z{\'u}{\~n}iga}, {Rosado}, {Ross}, {Rossi}, {Ruan},
		{Ruggeri}, {Rykoff}, {Salazar-Albornoz}, {Salvato}, {S{\'a}nchez}, {Aguado},
		{S{\'a}nchez-Gallego}, {Santana}, {Santiago}, {Sayres}, {Schiavon}, {da Silva
			Schimoia}, {Schlafly}, {Schlegel}, {Schneider}, {Schultheis}, {Schuster},
		{Schwope}, {Seo}, {Shao}, {Shen}, {Shetrone}, {Shull}, {Simon}, {Skinner},
		{Skrutskie}, {Slosar}, {Smith}, {Sobeck}, {Sobreira}, {Somers}, {Souto},
		{Stark}, {Stassun}, {Stauffer}, {Steinmetz}, {Storchi-Bergmann},
		{Streblyanska}, {Stringfellow}, {Su{\'a}rez}, {Sun}, {Suzuki}, {Szigeti},
		{Taghizadeh-Popp}, {Tang}, {Tao}, {Tayar}, {Tembe}, {Teske}, {Thakar},
		{Thomas}, {Thompson}, {Tinker}, {Tissera}, {Tojeiro}, {Hernandez Toledo}, {de
			la Torre}, {Tremonti}, {Troup}, {Valenzuela}, {Martinez Valpuesta},
		{Vargas-Gonz{\'a}lez}, {Vargas-Maga{\~n}a}, {Vazquez}, {Villanova}, {Vivek},
		{Vogt}, {Wake}, {Walterbos}, {Wang}, {Weaver}, {Weijmans}, {Weinberg},
		{Westfall}, {Whelan}, {Wild}, {Wilson}, {Wood-Vasey}, {Wylezalek}, {Xiao},
		{Yan}, {Yang}, {Ybarra}, {Y{\`e}che}, {Zakamska}, {Zamora}, {Zarrouk},
		{Zasowski}, {Zhang}, {Zhao}, {Zheng}, {Zheng}, {Zhou}, {Zhou}, {Zhu},
		{Zoccali}, \& {Zou}}]{Blanton2017}
	{Blanton}, M.~R., {Bershady}, M.~A., {Abolfathi}, B., {et~al.} 2017, \aj, 154,
	28
	
	\bibitem[{{Bowen} \& {Vaughan}(1973)}]{Bowen1973}
	{Bowen}, I.~S. \& {Vaughan}, A.~H., J. 1973, \ao, 12, 1430
	
	\bibitem[{{Bressan} {et~al.}(2012){Bressan}, {Marigo}, {Girardi}, {Salasnich},
		{Dal Cero}, {Rubele}, \& {Nanni}}]{2012MNRAS.427..127B}
	{Bressan}, A., {Marigo}, P., {Girardi}, L., {et~al.} 2012, \mnras, 427, 127
	
	\bibitem[{{Cardelli} {et~al.}(1989){Cardelli}, {Clayton}, \&
		{Mathis}}]{Cardelli1989}
	{Cardelli}, J.~A., {Clayton}, G.~C., \& {Mathis}, J.~S. 1989, \apj, 345, 245
	
	\bibitem[{{Carretta} \& {Bragaglia}(2021)}]{Carretta2021}
	{Carretta}, E. \& {Bragaglia}, A. 2021, \aap, 646, A9
	
	\bibitem[{{Carretta} {et~al.}(2010){Carretta}, {Bragaglia}, {Gratton},
		{Lucatello}, {Bellazzini}, \& {D'Orazi}}]{Carretta2010}
	{Carretta}, E., {Bragaglia}, A., {Gratton}, R., {et~al.} 2010, \apjl, 712, L21
	
	\bibitem[{{Carretta} {et~al.}(2009){Carretta}, {Bragaglia}, {Gratton},
		{Lucatello}, {Catanzaro}, {Leone}, {Bellazzini}, {Claudi}, {D'Orazi},
		{Momany}, {Ortolani}, {Pancino}, {Piotto}, {Recio-Blanco}, \&
		{Sabbi}}]{Carreta_2009A&A...505..117C}
	{Carretta}, E., {Bragaglia}, A., {Gratton}, R.~G., {et~al.} 2009, \aap, 505,
	117
	
	\bibitem[{{Cunha} {et~al.}(2017){Cunha}, {Smith}, {Hasselquist}, {Souto},
		{Shetrone}, {Allende Prieto}, {Bizyaev}, {Frinchaboy},
		{Garc{\'\i}a-Hern{\'a}ndez}, {Holtzman}, {Johnson}, {J{\H{o}}nsson},
		{Majewski}, {M{\'e}sz{\'a}ros}, {Nidever}, {Pinsonneault}, {Schiavon},
		{Sobeck}, {Skrutskie}, {Zamora}, {Zasowski}, \&
		{Fern{\'a}ndez-Trincado}}]{Cunha2017}
	{Cunha}, K., {Smith}, V.~V., {Hasselquist}, S., {et~al.} 2017, \apj, 844, 145
	
	\bibitem[{{D'Antona} {et~al.}(2016){D'Antona}, {Vesperini}, {D'Ercole},
		{Ventura}, {Milone}, {Marino}, \& {Tailo}}]{D'ANTONA_2016MNRAS.458.2122D}
	{D'Antona}, F., {Vesperini}, E., {D'Ercole}, A., {et~al.} 2016, \mnras, 458,
	2122
	
	\bibitem[{{de Mink} {et~al.}(2009){de Mink}, {Pols}, {Langer}, \&
		{Izzard}}]{deMink_2009A&A...507L...1D}
	{de Mink}, S.~E., {Pols}, O.~R., {Langer}, N., \& {Izzard}, R.~G. 2009, \aap,
	507, L1
	
	\bibitem[{{Decressin} {et~al.}(2007{\natexlab{a}}){Decressin}, {Meynet},
		{Charbonnel}, {Prantzos}, \& {Ekstr{\"o}m}}]{DECRESSIN_2007A&A...464.1029D}
	{Decressin}, T., {Meynet}, G., {Charbonnel}, C., {Prantzos}, N., \&
	{Ekstr{\"o}m}, S. 2007{\natexlab{a}}, \aap, 464, 1029
	
	\bibitem[{{Decressin} {et~al.}(2007{\natexlab{b}}){Decressin}, {Meynet},
		{Charbonnel}, {Prantzos}, \& {Ekstr{\"o}m}}]{Decressin2007}
	{Decressin}, T., {Meynet}, G., {Charbonnel}, C., {Prantzos}, N., \&
	{Ekstr{\"o}m}, S. 2007{\natexlab{b}}, \aap, 464, 1029
	
	\bibitem[{{Dutra-Ferreira} {et~al.}(2016){Dutra-Ferreira}, {Pasquini},
		{Smiljanic}, {Porto de Mello}, \& {Steffen}}]{Dutra2016}
	{Dutra-Ferreira}, L., {Pasquini}, L., {Smiljanic}, R., {Porto de Mello}, G.~F.,
	\& {Steffen}, M. 2016, \aap, 585, A75
	
	\bibitem[{{Fern{\'a}ndez-Trincado}
		{et~al.}(2021{\natexlab{a}}){Fern{\'a}ndez-Trincado}, {Beers}, {Barbuy},
		{M{\'e}sz{\'a}ros}, {Minniti}, {Smith}, {Cunha}, {Villanova}, {Geisler},
		{Majewski}, {Carigi}, {Tang}, {Moni Bidin}, \&
		{Vieira}}]{Fernandez-Trincado2021a}
	{Fern{\'a}ndez-Trincado}, J.~G., {Beers}, T.~C., {Barbuy}, B., {et~al.}
	2021{\natexlab{a}}, \apjl, 918, L9
	
	\bibitem[{{Fern{\'a}ndez-Trincado}
		{et~al.}(2022{\natexlab{a}}){Fern{\'a}ndez-Trincado}, {Beers}, {Barbuy},
		{Minniti}, {Chiappini}, {Garro}, {Tang}, {Alves-Brito}, {Villanova},
		{Geisler}, {Lane}, \& {Diaz}}]{Fernandez-Trincado2022Nitrogen}
	{Fern{\'a}ndez-Trincado}, J.~G., {Beers}, T.~C., {Barbuy}, B., {et~al.}
	2022{\natexlab{a}}, \aap, 663, A126
	
	\bibitem[{{Fern{\'a}ndez-Trincado}
		{et~al.}(2020{\natexlab{a}}){Fern{\'a}ndez-Trincado}, {Beers}, \&
		{Minniti}}]{Fernandez-Trincado2020a}
	{Fern{\'a}ndez-Trincado}, J.~G., {Beers}, T.~C., \& {Minniti}, D.
	2020{\natexlab{a}}, \aap, 644, A83
	
	\bibitem[{{Fern{\'a}ndez-Trincado}
		{et~al.}(2021{\natexlab{b}}){Fern{\'a}ndez-Trincado}, {Beers}, {Minniti},
		{Carigi}, {Placco}, {Chun}, {Lane}, {Geisler}, {Villanova}, {Souza},
		{Barbuy}, {P{\'e}rez-Villegas}, {Chiappini}, {Queiroz}, {Tang},
		{Alonso-Garc{\'\i}a}, {Piatti}, {Palma}, {Alves-Brito}, {Moni Bidin},
		{Roman-Lopes}, {Mu{\~n}oz}, {Singh}, {Kundu}, {Chaves-Velasquez},
		{Romero-Colmenares}, {Longa-Pe{\~n}a}, {Soto}, \&
		{Vieira}}]{Fernandez-Trincado2021b}
	{Fern{\'a}ndez-Trincado}, J.~G., {Beers}, T.~C., {Minniti}, D., {et~al.}
	2021{\natexlab{b}}, \aap, 647, A64
	
	\bibitem[{{Fern{\'a}ndez-Trincado}
		{et~al.}(2021{\natexlab{c}}){Fern{\'a}ndez-Trincado}, {Beers}, {Minniti},
		{Moni Bidin}, {Barbuy}, {Villanova}, {Geisler}, {Lane}, {Roman-Lopes}, \&
		{Bizyaev}}]{Fernandez-Trincado2021c}
	{Fern{\'a}ndez-Trincado}, J.~G., {Beers}, T.~C., {Minniti}, D., {et~al.}
	2021{\natexlab{c}}, \aap, 648, A70
	
	\bibitem[{{Fern{\'a}ndez-Trincado}
		{et~al.}(2020{\natexlab{b}}){Fern{\'a}ndez-Trincado}, {Minniti}, {Beers},
		{Villanova}, {Geisler}, {Souza}, {Smith}, {Placco}, {Vieira},
		{P{\'e}rez-Villegas}, {Barbuy}, {Alves-Brito}, {Bidin}, {Alonso-Garc{\'\i}a},
		{Tang}, \& {Palma}}]{Fernandez-Trincado2020b}
	{Fern{\'a}ndez-Trincado}, J.~G., {Minniti}, D., {Beers}, T.~C., {et~al.}
	2020{\natexlab{b}}, \aap, 643, A145
	
	\bibitem[{{Fern{\'a}ndez-Trincado}
		{et~al.}(2022{\natexlab{b}}){Fern{\'a}ndez-Trincado}, {Minniti}, {Garro}, \&
		{Villanova}}]{Fernandez-Trincado2022a}
	{Fern{\'a}ndez-Trincado}, J.~G., {Minniti}, D., {Garro}, E.~R., \& {Villanova},
	S. 2022{\natexlab{b}}, \aap, 657, A84
	
	\bibitem[{{Fern{\'a}ndez-Trincado}
		{et~al.}(2021{\natexlab{d}}){Fern{\'a}ndez-Trincado}, {Minniti}, {Souza},
		{Beers}, {Geisler}, {Moni Bidin}, {Villanova}, {Majewski}, {Barbuy},
		{P{\'e}rez-Villegas}, {Henao}, {Romero-Colmenares}, {Roman-Lopes}, \&
		{Lane}}]{Fernandez-Trincado2021d}
	{Fern{\'a}ndez-Trincado}, J.~G., {Minniti}, D., {Souza}, S.~O., {et~al.}
	2021{\natexlab{d}}, \apjl, 908, L42
	
	\bibitem[{{Fern{\'a}ndez-Trincado}
		{et~al.}(2022{\natexlab{c}}){Fern{\'a}ndez-Trincado}, {Villanova}, {Geisler},
		{Barbuy}, {Minniti}, {Beers}, {M{\'e}sz{\'a}ros}, {Tang}, {Cohen}, {Moni
			Bidin}, {Garro}, {Baeza}, \& {Mu{\~n}oz}}]{Fernandez-Trincado2022b}
	{Fern{\'a}ndez-Trincado}, J.~G., {Villanova}, S., {Geisler}, D., {et~al.}
	2022{\natexlab{c}}, \aap, 658, A116
	
	\bibitem[{{Fern{\'a}ndez-Trincado} {et~al.}(2019){Fern{\'a}ndez-Trincado},
		{Zamora}, {Souto}, {Cohen}, {Dell'Agli}, {Garc{\'\i}a-Hern{\'a}ndez},
		{Masseron}, {Schiavon}, {M{\'e}sz{\'a}ros}, {Cunha}, {Hasselquist},
		{Shetrone}, {Schiappacasse Ulloa}, {Tang}, {Geisler}, {Schleicher},
		{Villanova}, {Mennickent}, {Minniti}, {Alonso-Garc{\'\i}a}, {Manchado},
		{Beers}, {Sobeck}, {Zasowski}, {Schultheis}, {Majewski}, {Rojas-Arriagada},
		{Almeida}, {Santana}, {Oelkers}, {Longa-Pe{\~n}a}, {Carrera}, {Burgasser},
		{Lane}, {Roman-Lopes}, {Ivans}, \& {Hearty}}]{Fernandez-Trincado2019}
	{Fern{\'a}ndez-Trincado}, J.~G., {Zamora}, O., {Souto}, D., {et~al.} 2019,
	\aap, 627, A178
	
	\bibitem[{{Garc{\'\i}a P{\'e}rez} {et~al.}(2016){Garc{\'\i}a P{\'e}rez},
		{Allende Prieto}, {Holtzman}, {Shetrone}, {M{\'e}sz{\'a}ros}, {Bizyaev},
		{Carrera}, {Cunha}, {Garc{\'\i}a-Hern{\'a}ndez}, {Johnson}, {Majewski},
		{Nidever}, {Schiavon}, {Shane}, {Smith}, {Sobeck}, {Troup}, {Zamora},
		{Weinberg}, {Bovy}, {Eisenstein}, {Feuillet}, {Frinchaboy}, {Hayden},
		{Hearty}, {Nguyen}, {O'Connell}, {Pinsonneault}, {Wilson}, \&
		{Zasowski}}]{Garcia2016}
	{Garc{\'\i}a P{\'e}rez}, A.~E., {Allende Prieto}, C., {Holtzman}, J.~A.,
	{et~al.} 2016, \aj, 151, 144
	
	\bibitem[{{Geisler} {et~al.}(2021){Geisler}, {Villanova}, {O'Connell}, {Cohen},
		{Moni Bidin}, {Fern{\'a}ndez-Trincado}, {Mu{\~n}oz}, {Minniti}, {Zoccali},
		{Rojas-Arriagada}, {Contreras Ramos}, {Catelan}, {Mauro}, {Cort{\'e}s},
		{Ferreira Lopes}, {Arentsen}, {Starkenburg}, {Martin}, {Tang}, {Parisi},
		{Alonso-Garc{\'\i}a}, {Gran}, {Cunha}, {Smith}, {Majewski}, {J{\"o}nsson},
		{Garc{\'\i}a-Hern{\'a}ndez}, {Horta}, {M{\'e}sz{\'a}ros}, {Monaco},
		{Monachesi}, {Mu{\~n}oz}, {Brownstein}, {Beers}, {Lane}, {Barbuy}, {Sobeck},
		{Henao}, {Gonz{\'a}lez-D{\'\i}az}, {Miranda}, {Reinarz}, \&
		{Santander}}]{Geisler_2021_2021A&A...652A.157G}
	{Geisler}, D., {Villanova}, S., {O'Connell}, J.~E., {et~al.} 2021, \aap, 652,
	A157
	
	\bibitem[{{Gonz{\'a}lez-D{\'\i}az} {et~al.}(2023){Gonz{\'a}lez-D{\'\i}az},
		{Fern{\'a}ndez-Trincado}, {Villanova}, {Geisler}, {Barbuy}, {Minniti},
		{Beers}, {Moni Bidin}, {Mauro}, {Mu{\~n}oz}, {Tang}, {Soto}, {Monachesi},
		{Lane}, \& {Frelijj}}]{2023NGC6558}
	{Gonz{\'a}lez-D{\'\i}az}, D., {Fern{\'a}ndez-Trincado}, J.~G., {Villanova}, S.,
	{et~al.} 2023, \mnras, 526, 6274
	
	\bibitem[{{Grevesse} {et~al.}(2015){Grevesse}, {Scott}, {Asplund}, \&
		{Sauval}}]{Grevesse2015}
	{Grevesse}, N., {Scott}, P., {Asplund}, M., \& {Sauval}, A.~J. 2015, \aap, 573,
	A27
	
	\bibitem[{{Gustafsson} {et~al.}(2008){Gustafsson}, {Edvardsson}, {Eriksson},
		{J{\o}rgensen}, {Nordlund}, \& {Plez}}]{Gustafsson2008}
	{Gustafsson}, B., {Edvardsson}, B., {Eriksson}, K., {et~al.} 2008, \aap, 486,
	951
	
	\bibitem[{{Harris}(1996)}]{Harris_1996AJ....112.1487H}
	{Harris}, W.~E. 1996, \aj, 112, 1487
	
	\bibitem[{{Harris}(2010)}]{Harris_2010arXiv1012.3224H}
	{Harris}, W.~E. 2010, arXiv e-prints, arXiv:1012.3224
	
	\bibitem[{{Hasselquist} {et~al.}(2016){Hasselquist}, {Shetrone}, {Cunha},
		{Smith}, {Holtzman}, {Lawler}, {Allende Prieto}, {Beers}, {Chojnowski},
		{Fern{\'a}ndez-Trincado}, {Garc{\'\i}a-Hern{\'a}ndez}, {Hearty}, {Majewski},
		{Pereira}, {Placco}, {Villanova}, \& {Zamora}}]{Hasselquist2016}
	{Hasselquist}, S., {Shetrone}, M., {Cunha}, K., {et~al.} 2016, \apj, 833, 81
	
	\bibitem[{{Holtzman} {et~al.}(2018){Holtzman}, {Hasselquist}, {Shetrone},
		{Cunha}, {Allende Prieto}, {Anguiano}, {Bizyaev}, {Bovy}, {Casey},
		{Edvardsson}, {Johnson}, {J{\"o}nsson}, {Meszaros}, {Smith}, {Sobeck},
		{Zamora}, {Chojnowski}, {Fernandez-Trincado}, {Garcia-Hernandez}, {Majewski},
		{Pinsonneault}, {Souto}, {Stringfellow}, {Tayar}, {Troup}, \&
		{Zasowski}}]{Holtzman2018}
	{Holtzman}, J.~A., {Hasselquist}, S., {Shetrone}, M., {et~al.} 2018, \aj, 156,
	125
	
	\bibitem[{{J{\"o}nsson} {et~al.}(2018){J{\"o}nsson}, {Allende Prieto},
		{Holtzman}, {Feuillet}, {Hawkins}, {Cunha}, {M{\'e}sz{\'a}ros},
		{Hasselquist}, {Fern{\'a}ndez-Trincado}, {Garc{\'\i}a-Hern{\'a}ndez},
		{Bizyaev}, {Carrera}, {Majewski}, {Pinsonneault}, {Shetrone}, {Smith},
		{Sobeck}, {Souto}, {Stringfellow}, {Teske}, \& {Zamora}}]{Jonsson_2018}
	{J{\"o}nsson}, H., {Allende Prieto}, C., {Holtzman}, J.~A., {et~al.} 2018, \aj,
	156, 126
	
	\bibitem[{{Kerber} {et~al.}(2019){Kerber}, {Libralato}, {Souza}, {Oliveira},
		{Ortolani}, {P{\'e}rez-Villegas}, {Barbuy}, {Dias}, {Bica}, \&
		{Nardiello}}]{Kerber2019}
	{Kerber}, L.~O., {Libralato}, M., {Souza}, S.~O., {et~al.} 2019, \mnras, 484,
	5530
	
	\bibitem[{{Majewski} {et~al.}(2017){Majewski}, {Schiavon}, {Frinchaboy},
		{Allende Prieto}, {Barkhouser}, {Bizyaev}, {Blank}, {Brunner}, {Burton},
		{Carrera}, {Chojnowski}, {Cunha}, {Epstein}, {Fitzgerald}, {Garc{\'\i}a
			P{\'e}rez}, {Hearty}, {Henderson}, {Holtzman}, {Johnson}, {Lam}, {Lawler},
		{Maseman}, {M{\'e}sz{\'a}ros}, {Nelson}, {Nguyen}, {Nidever}, {Pinsonneault},
		{Shetrone}, {Smee}, {Smith}, {Stolberg}, {Skrutskie}, {Walker}, {Wilson},
		{Zasowski}, {Anders}, {Basu}, {Beland}, {Blanton}, {Bovy}, {Brownstein},
		{Carlberg}, {Chaplin}, {Chiappini}, {Eisenstein}, {Elsworth}, {Feuillet},
		{Fleming}, {Galbraith-Frew}, {Garc{\'\i}a}, {Garc{\'\i}a-Hern{\'a}ndez},
		{Gillespie}, {Girardi}, {Gunn}, {Hasselquist}, {Hayden}, {Hekker}, {Ivans},
		{Kinemuchi}, {Klaene}, {Mahadevan}, {Mathur}, {Mosser}, {Muna}, {Munn},
		{Nichol}, {O'Connell}, {Parejko}, {Robin}, {Rocha-Pinto}, {Schultheis},
		{Serenelli}, {Shane}, {Silva Aguirre}, {Sobeck}, {Thompson}, {Troup},
		{Weinberg}, \& {Zamora}}]{Majewski_2017AJ....154...94M}
	{Majewski}, S.~R., {Schiavon}, R.~P., {Frinchaboy}, P.~M., {et~al.} 2017, \aj,
	154, 94
	
	\bibitem[{{Massari} {et~al.}(2019){Massari}, {Koppelman}, \&
		{Helmi}}]{Massari2019}
	{Massari}, D., {Koppelman}, H.~H., \& {Helmi}, A. 2019, \aap, 630, L4
	
	\bibitem[{{Masseron} {et~al.}(2019){Masseron}, {Garc{\'\i}a-Hern{\'a}ndez},
		{M{\'e}sz{\'a}ros}, {Zamora}, {Dell'Agli}, {Allende Prieto}, {Edvardsson},
		{Shetrone}, {Plez}, {Fern{\'a}ndez-Trincado}, {Cunha}, {J{\"o}nsson},
		{Geisler}, {Beers}, \& {Cohen}}]{Masseron2019}
	{Masseron}, T., {Garc{\'\i}a-Hern{\'a}ndez}, D.~A., {M{\'e}sz{\'a}ros}, S.,
	{et~al.} 2019, \aap, 622, A191
	
	\bibitem[{{Masseron} {et~al.}(2016){Masseron}, {Merle}, \&
		{Hawkins}}]{Masseron2016}
	{Masseron}, T., {Merle}, T., \& {Hawkins}, K. 2016, {BACCHUS: Brussels
		Automatic Code for Characterizing High accUracy Spectra}
	
	\bibitem[{{M{\'e}sz{\'a}ros} {et~al.}(2015){M{\'e}sz{\'a}ros}, {Martell},
		{Shetrone}, {Lucatello}, {Troup}, {Bovy}, {Cunha},
		{Garc{\'\i}a-Hern{\'a}ndez}, {Overbeek}, {Allende Prieto}, {Beers},
		{Frinchaboy}, {Garc{\'\i}a P{\'e}rez}, {Hearty}, {Holtzman}, {Majewski},
		{Nidever}, {Schiavon}, {Schneider}, {Sobeck}, {Smith}, {Zamora}, \&
		{Zasowski}}]{Meszaros2015}
	{M{\'e}sz{\'a}ros}, S., {Martell}, S.~L., {Shetrone}, M., {et~al.} 2015, \aj,
	149, 153
	
	\bibitem[{{M{\'e}sz{\'a}ros} {et~al.}(2021){M{\'e}sz{\'a}ros}, {Masseron},
		{Fern{\'a}ndez-Trincado}, {Garc{\'\i}a-Hern{\'a}ndez}, {Szigeti}, {Cunha},
		{Shetrone}, {Smith}, {Beaton}, {Beers}, {Brownstein}, {Geisler}, {Hayes},
		{J{\"o}nsson}, {Lane}, {Majewski}, {Minniti}, {Munoz}, {Nitschelm},
		{Roman-Lopes}, \& {Zamora}}]{Meszaros_2021MNRAS.505.1645M}
	{M{\'e}sz{\'a}ros}, S., {Masseron}, T., {Fern{\'a}ndez-Trincado}, J.~G.,
	{et~al.} 2021, \mnras, 505, 1645
	
	\bibitem[{{M{\'e}sz{\'a}ros} {et~al.}(2020){M{\'e}sz{\'a}ros}, {Masseron},
		{Garc{\'\i}a-Hern{\'a}ndez}, {Allende Prieto}, {Beers}, {Bizyaev},
		{Chojnowski}, {Cohen}, {Cunha}, {Dell'Agli}, {Ebelke},
		{Fern{\'a}ndez-Trincado}, {Frinchaboy}, {Geisler}, {Hasselquist}, {Hearty},
		{Holtzman}, {Johnson}, {Lane}, {Lacerna}, {Longa-Pe{\~n}a}, {Majewski},
		{Martell}, {Minniti}, {Nataf}, {Nidever}, {Pan}, {Schiavon}, {Shetrone},
		{Smith}, {Sobeck}, {Stringfellow}, {Szigeti}, {Tang}, {Wilson}, \&
		{Zamora}}]{Meszaros_2020}
	{M{\'e}sz{\'a}ros}, S., {Masseron}, T., {Garc{\'\i}a-Hern{\'a}ndez}, D.~A.,
	{et~al.} 2020, \mnras, 492, 1641
	
	\bibitem[{{Minniti}(1995)}]{Minniti1995}
	{Minniti}, D. 1995, \aj, 109, 1663
	
	\bibitem[{{Minniti} {et~al.}(2017{\natexlab{a}}){Minniti},
		{Alonso-Garc{\'\i}a}, {Braga}, {Contreras Ramos}, {Hempel}, {Palma},
		{Pullen}, \& {Saito}}]{Minniti_a_2017RNAAS...1...16M}
	{Minniti}, D., {Alonso-Garc{\'\i}a}, J., {Braga}, V., {et~al.}
	2017{\natexlab{a}}, Research Notes of the American Astronomical Society, 1,
	16
	
	\bibitem[{{Minniti} {et~al.}(2017{\natexlab{b}}){Minniti}, {Geisler},
		{Alonso-Garc{\'\i}a}, {Palma}, {Beam{\'\i}n}, {Borissova}, {Catelan},
		{Clari{\'a}}, {Cohen}, {Contreras Ramos}, {Dias}, {Fern{\'a}ndez-Trincado},
		{G{\'o}mez}, {Hempel}, {Ivanov}, {Kurtev}, {Lucas}, {Moni-Bidin}, {Pullen},
		{Ram{\'\i}rez Alegr{\'\i}a}, {Saito}, \&
		{Valenti}}]{Minniti_b_2017ApJ...849L..24M}
	{Minniti}, D., {Geisler}, D., {Alonso-Garc{\'\i}a}, J., {et~al.}
	2017{\natexlab{b}}, \apjl, 849, L24
	
	\bibitem[{{Nataf} {et~al.}(2019){Nataf}, {Wyse}, {Schiavon}, {Ting}, {Minniti},
		{Cohen}, {Fern{\'a}ndez-Trincado}, {Geisler}, {Nitschelm}, \&
		{Frinchaboy}}]{Nataf2019}
	{Nataf}, D.~M., {Wyse}, R. F.~G., {Schiavon}, R.~P., {et~al.} 2019, \aj, 158,
	14
	
	\bibitem[{{Nidever} {et~al.}(2015){Nidever}, {Holtzman}, {Allende Prieto},
		{Beland}, {Bender}, {Bizyaev}, {Burton}, {Desphande}, {Fleming}, {Garc{\'\i}a
			P{\'e}rez}, {Hearty}, {Majewski}, {M{\'e}sz{\'a}ros}, {Muna}, {Nguyen},
		{Schiavon}, {Shetrone}, {Skrutskie}, {Sobeck}, \& {Wilson}}]{Nidever2015}
	{Nidever}, D.~L., {Holtzman}, J.~A., {Allende Prieto}, C., {et~al.} 2015, \aj,
	150, 173
	
	\bibitem[{{Palma} {et~al.}(2019){Palma}, {Minniti}, {Alonso-Garc{\'\i}a},
		{Crestani}, {Netzel}, {Clari{\'a}}, {Saito}, {Dias},
		{Fern{\'a}ndez-Trincado}, {Kammers}, {Geisler}, {G{\'o}mez}, {Hempel}, \&
		{Pullen}}]{Palma_2019MNRAS.487.3140P}
	{Palma}, T., {Minniti}, D., {Alonso-Garc{\'\i}a}, J., {et~al.} 2019, \mnras,
	487, 3140
	
	\bibitem[{{Pancino} {et~al.}(2017){Pancino}, {Romano}, {Tang},
		{Tautvai{\v{s}}ien{\.{e}}}, {Casey}, {Gruyters}, {Geisler}, {San Roman},
		{Randich}, {Alfaro}, {Bragaglia}, {Flaccomio}, {Korn}, {Recio-Blanco},
		{Smiljanic}, {Carraro}, {Bayo}, {Costado}, {Damiani}, {Jofr{\'e}}, {Lardo},
		{de Laverny}, {Monaco}, {Morbidelli}, {Sbordone}, {Sousa}, \&
		{Villanova}}]{Pancino2017}
	{Pancino}, E., {Romano}, D., {Tang}, B., {et~al.} 2017, \aap, 601, A112
	
	\bibitem[{{Plez}(2012)}]{Plez2012}
	{Plez}, B. 2012, {Turbospectrum: Code for spectral synthesis}
	
	\bibitem[{{Renzini} {et~al.}(2015){Renzini}, {D'Antona}, {Cassisi}, {King},
		{Milone}, {Ventura}, {Anderson}, {Bedin}, {Bellini}, {Brown}, {Piotto}, {van
			der Marel}, {Barbuy}, {Dalessandro}, {Hidalgo}, {Marino}, {Ortolani},
		{Salaris}, \& {Sarajedini}}]{Renzini_2015MNRAS.454.4197R}
	{Renzini}, A., {D'Antona}, F., {Cassisi}, S., {et~al.} 2015, \mnras, 454, 4197
	
	\bibitem[{{Romero-Colmenares} {et~al.}(2021){Romero-Colmenares},
		{Fern{\'a}ndez-Trincado}, {Geisler}, {Souza}, {Villanova}, {Longa-Pe{\~n}a},
		{Minniti}, {Beers}, {Bidin}, {Perez-Villegas}, {Moreno}, {Garro}, {Baeza},
		{Henao}, {Barbuy}, {Alonso-Garc{\'\i}a}, {Cohen}, {Lane}, \&
		{Mu{\~n}oz}}]{Romero_2021A&A...652A.158R}
	{Romero-Colmenares}, M., {Fern{\'a}ndez-Trincado}, J.~G., {Geisler}, D.,
	{et~al.} 2021, \aap, 652, A158
	
	\bibitem[{{Saito} {et~al.}(2012){Saito}, {Hempel}, {Minniti}, {Lucas},
		{Rejkuba}, {Toledo}, {Gonzalez}, {Alonso-Garc{\'\i}a}, {Irwin},
		{Gonzalez-Solares}, {Hodgkin}, {Lewis}, {Cross}, {Ivanov}, {Kerins},
		{Emerson}, {Soto}, {Am{\^o}res}, {Gurovich}, {D{\'e}k{\'a}ny}, {Angeloni},
		{Beamin}, {Catelan}, {Padilla}, {Zoccali}, {Pietrukowicz}, {Moni Bidin},
		{Mauro}, {Geisler}, {Folkes}, {Sale}, {Borissova}, {Kurtev}, {Ahumada},
		{Alonso}, {Adamson}, {Arias}, {Bandyopadhyay}, {Barb{\'a}}, {Barbuy},
		{Baume}, {Bedin}, {Bellini}, {Benjamin}, {Bica}, {Bonatto}, {Bronfman},
		{Carraro}, {Chen{\`e}}, {Clari{\'a}}, {Clarke}, {Contreras}, {Corvill{\'o}n},
		{de Grijs}, {Dias}, {Drew}, {Fari{\~n}a}, {Feinstein},
		{Fern{\'a}ndez-Laj{\'u}s}, {Gamen}, {Gieren}, {Goldman},
		{Gonz{\'a}lez-Fern{\'a}ndez}, {Grand}, {Gunthardt}, {Hambly}, {Hanson},
		{He{\l}miniak}, {Hoare}, {Huckvale}, {Jord{\'a}n}, {Kinemuchi}, {Longmore},
		{L{\'o}pez-Corredoira}, {Maccarone}, {Majaess}, {Mart{\'\i}n}, {Masetti},
		{Mennickent}, {Mirabel}, {Monaco}, {Morelli}, {Motta}, {Palma}, {Parisi},
		{Parker}, {Pe{\~n}aloza}, {Pietrzy{\'n}ski}, {Pignata}, {Popescu}, {Read},
		{Rojas}, {Roman-Lopes}, {Ruiz}, {Saviane}, {Schreiber}, {Schr{\"o}der},
		{Sharma}, {Smith}, {Sodr{\'e}}, {Stead}, {Stephens}, {Tamura}, {Tappert},
		{Thompson}, {Valenti}, {Vanzi}, {Walton}, {Weidmann}, \&
		{Zijlstra}}]{Saito_2012_hp1_vvv_2012A&A...537A.107S}
	{Saito}, R.~K., {Hempel}, M., {Minniti}, D., {et~al.} 2012, \aap, 537, A107
	
	\bibitem[{{Schiavon} {et~al.}(2017{\natexlab{a}}){Schiavon}, {Johnson},
		{Frinchaboy}, {Zasowski}, {M{\'e}sz{\'a}ros}, {Garc{\'\i}a-Hern{\'a}ndez},
		{Cohen}, {Tang}, {Villanova}, {Geisler}, {Beers}, {Fern{\'a}ndez-Trincado},
		{Garc{\'\i}a P{\'e}rez}, {Lucatello}, {Majewski}, {Martell}, {O'Connell},
		{Allende Prieto}, {Bizyaev}, {Carrera}, {Lane}, {Malanushenko},
		{Malanushenko}, {Mu{\~n}oz}, {Nitschelm}, {Oravetz}, {Pan}, {Roman-Lopes},
		{Schultheis}, \& {Simmons}}]{Schiavon2017}
	{Schiavon}, R.~P., {Johnson}, J.~A., {Frinchaboy}, P.~M., {et~al.}
	2017{\natexlab{a}}, \mnras, 466, 1010
	
	\bibitem[{{Schiavon} {et~al.}(2024){Schiavon}, {Phillips}, {Myers}, {Horta},
		{Minniti}, {Allende Prieto}, {Anguiano}, {Beaton}, {Beers}, {Brownstein},
		{Cohen}, {Fern{\'a}ndez-Trincado}, {Frinchaboy}, {J{\"o}nsson}, {Kisku},
		{Lane}, {Majewski}, {Mason}, {M{\'e}sz{\'a}ros}, \&
		{Stringfellow}}]{Schiavon2024}
	{Schiavon}, R.~P., {Phillips}, S.~G., {Myers}, N., {et~al.} 2024, \mnras, 528,
	1393
	
	\bibitem[{{Schiavon} {et~al.}(2017{\natexlab{b}}){Schiavon}, {Zamora},
		{Carrera}, {Lucatello}, {Robin}, {Ness}, {Martell}, {Smith},
		{Garc{\'\i}a-Hern{\'a}ndez}, {Manchado}, {Sch{\"o}nrich}, {Bastian},
		{Chiappini}, {Shetrone}, {Mackereth}, {Williams}, {M{\'e}sz{\'a}ros},
		{Allende Prieto}, {Anders}, {Bizyaev}, {Beers}, {Chojnowski}, {Cunha},
		{Epstein}, {Frinchaboy}, {Garc{\'\i}a P{\'e}rez}, {Hearty}, {Holtzman},
		{Johnson}, {Kinemuchi}, {Majewski}, {Muna}, {Nidever}, {Nguyen}, {O'Connell},
		{Oravetz}, {Pan}, {Pinsonneault}, {Schneider}, {Schultheis}, {Simmons},
		{Skrutskie}, {Sobeck}, {Wilson}, \& {Zasowski}}]{Schiavon2017Nitrogen}
	{Schiavon}, R.~P., {Zamora}, O., {Carrera}, R., {et~al.} 2017{\natexlab{b}},
	\mnras, 465, 501
	
	\bibitem[{{Shetrone} {et~al.}(2015){Shetrone}, {Bizyaev}, {Lawler}, {Allende
			Prieto}, {Johnson}, {Smith}, {Cunha}, {Holtzman}, {Garc{\'\i}a P{\'e}rez},
		{M{\'e}sz{\'a}ros}, {Sobeck}, {Zamora}, {Garc{\'\i}a-Hern{\'a}ndez}, {Souto},
		{Chojnowski}, {Koesterke}, {Majewski}, \& {Zasowski}}]{Shetrone2015}
	{Shetrone}, M., {Bizyaev}, D., {Lawler}, J.~E., {et~al.} 2015, \apjs, 221, 24
	
	\bibitem[{{Simpson} \& {Martell}(2019)}]{Simpson_2019MNRAS.490..741S}
	{Simpson}, J.~D. \& {Martell}, S.~L. 2019, \mnras, 490, 741
	
	\bibitem[{{Smith} {et~al.}(2021){Smith}, {Bizyaev}, {Cunha}, {Shetrone},
		{Souto}, {Allende Prieto}, {Masseron}, {M{\'e}sz{\'a}ros}, {J{\"o}nsson},
		{Hasselquist}, {Osorio}, {Garc{\'\i}a-Hern{\'a}ndez}, {Plez}, {Beaton},
		{Holtzman}, {Majewski}, {Stringfellow}, \& {Sobeck}}]{Smith2021}
	{Smith}, V.~V., {Bizyaev}, D., {Cunha}, K., {et~al.} 2021, \aj, 161, 254
	
	\bibitem[{{Smith} {et~al.}(2013){Smith}, {Cunha}, {Shetrone}, {Meszaros},
		{Allende Prieto}, {Bizyaev}, {Garc{\'\i}a P{\'e}rez}, {Majewski}, {Schiavon},
		{Holtzman}, \& {Johnson}}]{Smith2013}
	{Smith}, V.~V., {Cunha}, K., {Shetrone}, M.~D., {et~al.} 2013, \apj, 765, 16
	
	\bibitem[{{Vasiliev} \& {Baumgardt}(2021)}]{Baumgardt_2021}
	{Vasiliev}, E. \& {Baumgardt}, H. 2021, \mnras, 505, 5978
	
	\bibitem[{{Ventura} {et~al.}(2016){Ventura}, {Garc{\'\i}a-Hern{\'a}ndez},
		{Dell'Agli}, {D'Antona}, {M{\'e}sz{\'a}ros}, {Lucatello}, {Di Criscienzo},
		{Shetrone}, {Tailo}, {Tang}, \& {Zamora}}]{Ventura2016}
	{Ventura}, P., {Garc{\'\i}a-Hern{\'a}ndez}, D.~A., {Dell'Agli}, F., {et~al.}
	2016, \apjl, 831, L17
	
	\bibitem[{{Wiescher} {et~al.}(2010){Wiescher}, {G{\"o}rres}, {Uberseder},
		{Imbriani}, \& {Pignatari}}]{Wiescher2010}
	{Wiescher}, M., {G{\"o}rres}, J., {Uberseder}, E., {Imbriani}, G., \&
	{Pignatari}, M. 2010, Annual Review of Nuclear and Particle Science, 60, 381
	
	\bibitem[{{Zamora} {et~al.}(2015){Zamora}, {Garc{\'\i}a-Hern{\'a}ndez},
		{Allende Prieto}, {Carrera}, {Koesterke}, {Edvardsson}, {Castelli}, {Plez},
		{Bizyaev}, {Cunha}, {Garc{\'\i}a P{\'e}rez}, {Gustafsson}, {Holtzman},
		{Lawler}, {Majewski}, {Manchado}, {M{\'e}sz{\'a}ros}, {Shane}, {Shetrone},
		{Smith}, \& {Zasowski}}]{Zamora2015}
	{Zamora}, O., {Garc{\'\i}a-Hern{\'a}ndez}, D.~A., {Allende Prieto}, C.,
	{et~al.} 2015, \aj, 149, 181
	
\end{thebibliography}

\end{document}